# REPRODUCTIVE SYSTEM AND INTERACTION WITH FAUNA IN A MEDITERRANEAN PYROPHITE SHRUB

*Josué-Antonio Nescolarde-Selva\*, José-Luis Usó-Doménech, Kristian Alonso-Stenberg*

*Department of Applied Mathematics. University of Alicante. Alicante. Spain.*

*\*Corresponding author: email: josue.selva@ua.es, Telephone. 0034680418381, Fax number. 0034 965 909 707*

**ABSTRACT**

The ULEX model, in its present state, involves the study of the biomass and the population of the shrub *Ulex parviflorus Pourret*, but while being a dynamic model, it is static in the sense that it does not imply the appearance of new specimens of this plant. As a complement to the ULEX model in its two dynamic and spatial aspects, and with the idea of extending the model, the authors have introduced from a biological and statistical point of view four characteristics of this species, flowering, pollination, fructification, taking special interest in the role played by the pollinators (bees) and dispersion of seeds.

## 1. INTRODUCTION

Gorse (*Ulex parviflorus Pourret*) is a Fabaceae (family *Fabaceae*). There are about 20 species native to the European Atlantic territory and the western part of the Mediterranean region (Bonet and Pausas, 2004). The *Ulex parviflorus Pourret* is a pyrophite species, and is the first colonizer in post-fire stages of ecological regeneration (Figure 1).

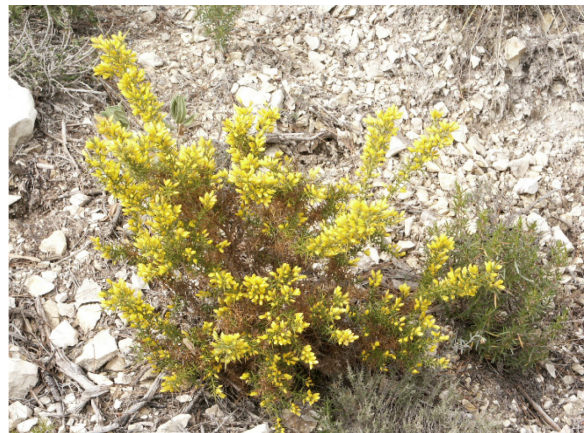

**Figure 1: The *Ulex parviflorus P.***

The fundamental role of the species *Ulex parviflorus Pourret* is manifested in its close relationship with forest fires, and together with the *Pinus halepensis* is the most important pyrophite species of the Mediterranean forest (Baeza, Raventós, Escarré and Vallejo, 2006; De Luis, Baeza, Raventós, González-Hidalgo, 2004; Dimitrakopoulos and Panov, 2001; Papió and Trabaud , 1990, 1991; Pausas and Verdú, 2008; Pausas et al., 2016;

Ripley et al., 2015; García et al., 2016; Hernández et al., 2010; Delitti et al., 2005; Santana et al, 2014; Cruz et al., 2003).

This pyrophite character is evident in:

1) Increased level of flammable material.
2) Germination of seeds that are favoured by the fires and the lack of competition in relation to other species which are removed by fire.

A mathematical population and reproductive model (ULEX model) of this species has been developed by Usó-Doménech, Nescolarde-Selva, Lloret-Climent and González-Franco (2018) and Usó-Doménech, Nescolarde-Selva, Lloret-Climent, González-Franco and Alonso-Stenberg, (2018).

On the other hand, we know that the ULEX dynamic model only involves the study of the biomass and population of the *Ulex parviflorus Pourret* shrub. Likewise, the model presents static characteristics, since it does not consider the appearance of new specimens of this shrub. Therefore, it is fundamental to promote this model from a dynamic point of view.

In this sense, through this paper we intend to extend the ULEX model through 4 characteristics of such plant, considering both the biological and statistical aspects: (1) flowering, (2) pollination (with special emphasis on bees as pollinators), (3) fructification and (4) seed dispersal (taking into account the role of ants). We carried out this study on specimens of *Ulex parviflorus Pourret*, located on a 100 $m^2$ plot in the *Desert de les Palmes* (*Castellón*, Spain).

## 2. THE STUDY AREA

The plot studied are located within the province of Castellón (Comunidad Valenciana, Spain). The experimentation plot is in the *Desert de les Palmes* (*Monte Bartolo*) (Figure 2).

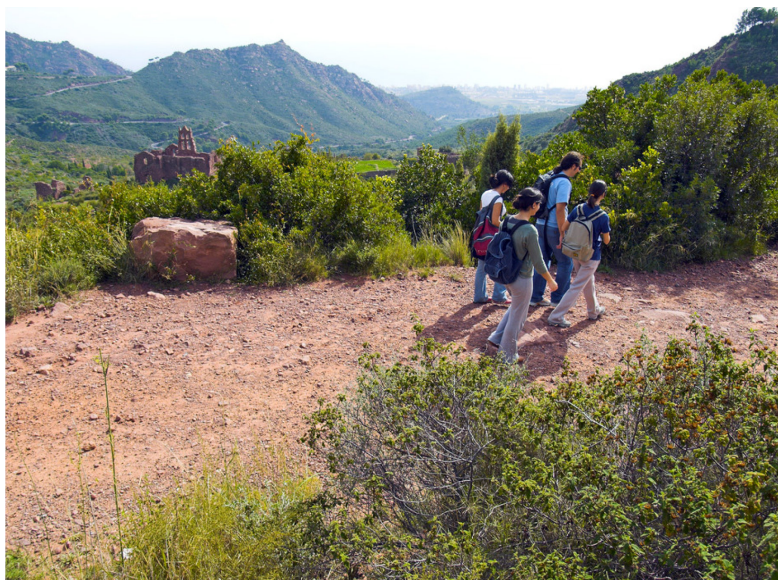

**Figure 2: Experimentation site.**

The plot studied is located within the province of Castellón (Spain) in the Desert de les Palmes (Monte Bartolo). This area has been chosen due to its lithology, soils, topography, climate, etc., considered representative of the Mediterranean terrestrial ecosystem. The plot is 100 m2 and an attempt has been made to keep the number of plants around Ulex parviflorus as low as possible in order to better carry out sampling and reduce interferences. All the specimens present inside the plot were placed on a map (Table 1).

**TABLE 1**
**Characteristics of the plot**

|  | **Plot of experimentation** |
|---|---|
| Location | Desert de les Palmes |
| UTM | 31 T BE 4741 |
| Altitude (m) | 550 |
| Reforestation | Yes |
| Climatic characteristics[1] (*) | Average annual temperature:15,2 ºC<br>Average temperature of the coldest month (January): 3º C<br>Average temperature of the warmest month (July): 29,4 ºC<br>Average annual precipitation:540 mm |
| List of plants | *Ulex parviflorus* Pourret<br>*Cistus monspeliensis* L.<br>*Cistus albidus* L.<br>*Pistacia lentiscus* L.<br>*Quercus coccifera* L.<br>*Juniperus oxycedrus* L. *subsp. oxycedrus*<br>*Pinus halepensis* Miller<br>*Rosmarinus officinalis* L.<br>*Chamaerops humilis* L.<br>*Inula viscosa* L. (Aiton) |

(*)The climatic characteristics, have been obtained on the basis of 33 years of observations (1941-1974). Source of data: Quereda (1976, 1985).

---

[1] Since September 1991 an Automatic Meteorological Station is in operation on top of Monte Bartolo (735 meters above sea level and 6 km from the sea) forming part of the network of observatories of the Universitat Jaume I. Its function is: to record of all climatic and environmental variables that it is equipped to measure automatically (including solar radiation, infrared radiation, soil and air temperature, atmospheric pressure, winds in direction and force, relative humidity, rainfall, etc.) These are the main exogenous variables taken into account in the model. All registers correspond to analogue inputs in 16 channels, with continuous scanning every two seconds, which can be processed, in any interval chosen from 5 minutes, using averages, maxima and minima, as well as filters of alternating values. It also has a digital input line for events such as rain and relative humidity. Data is registered on a hard drive using a duplex RS232C and data can be downloaded in ASCII format.

## 3. FLOWERING

### 3.1. Characteristics of the flower
The flowers are typically of papillary morphology, with a length of 11-14 mm. arranged in fascicles on primary and secondary spines of the branches produced the previous year. The inflorescence appears in clusters of three or more flowers located at the apex of the branches and sometimes on the spines (Figure 3).

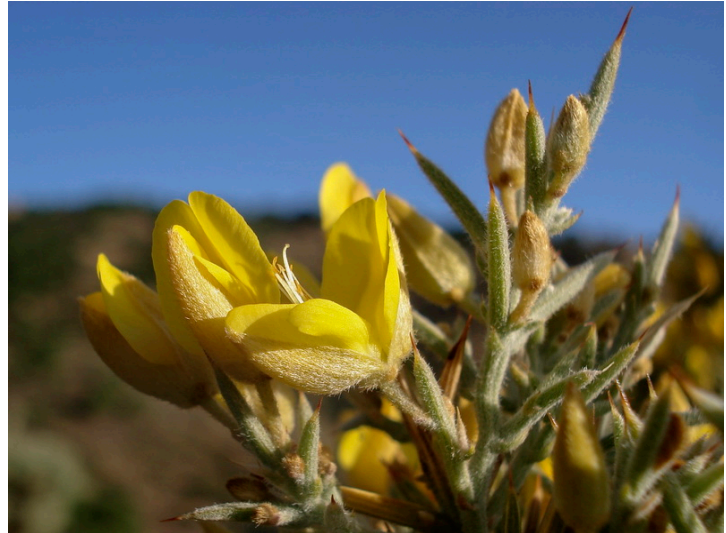

**Figure 3: Flowers of *Ulex parviflorus*.**
**Photo by courtesy of https://flora.biologiasur.org**

Flowers are hermaphrodite and do not produce nectar so pollen is the only reward offered to insects (Baeza, 2001; Carrión et al., 2001). Pollination is entomophilous, mainly produced by species of the genus *Apis* (Herrera, 1985).

### 3.2. Study of flowering.
It is a winter flowering plant, it develops from mid-autumn to late winter. Up to 80 days with intense flowering has been observed in the same individual (Herrera, 1985). In spring after flowering, the main branches begin to grow forming new branches on the branches of the previous year. Fieldwork has been carried out for three consecutive years, distributed as follows, the first and second year of sampling, from October to July of the first year. From October of the first year until July of the second year, the flowers were counted during the whole period of flowering. In the third year, the evolution from the period of maximum flowering to its disappearance was studied, from February to July.
The study of the period of flowering of this species from October to July, shows that the period of maximum flowering corresponds to February and March (Figures: 4, 5 and 6).

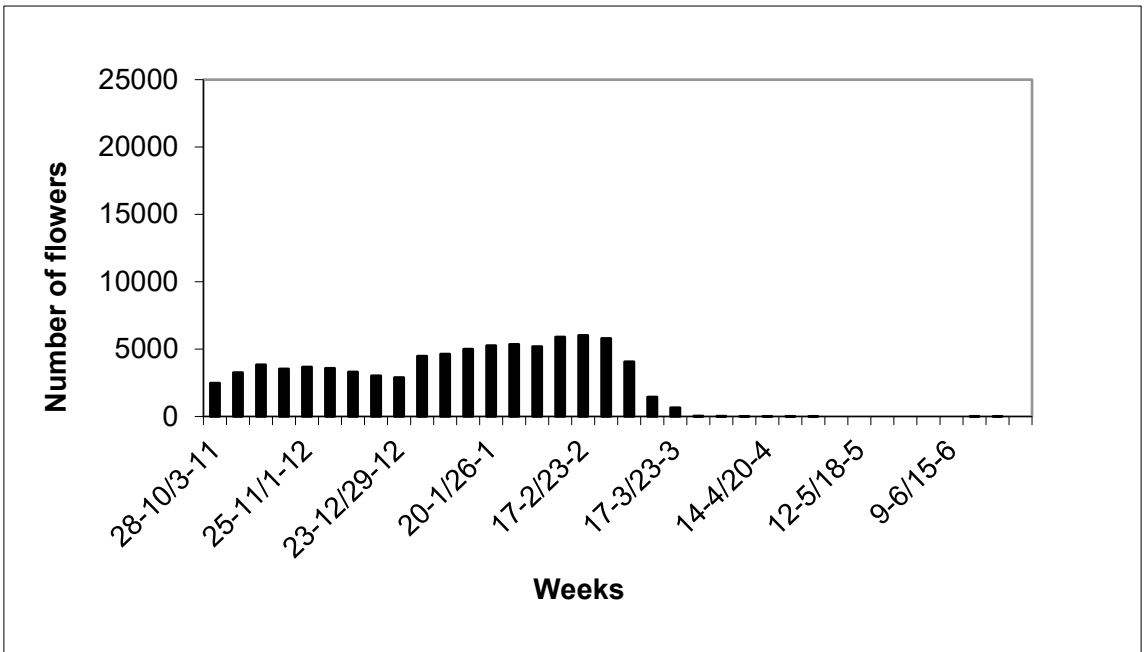

**Figure 4:** *Ulex parviflorus* flowers. From 28-October to 6-July of the first year.

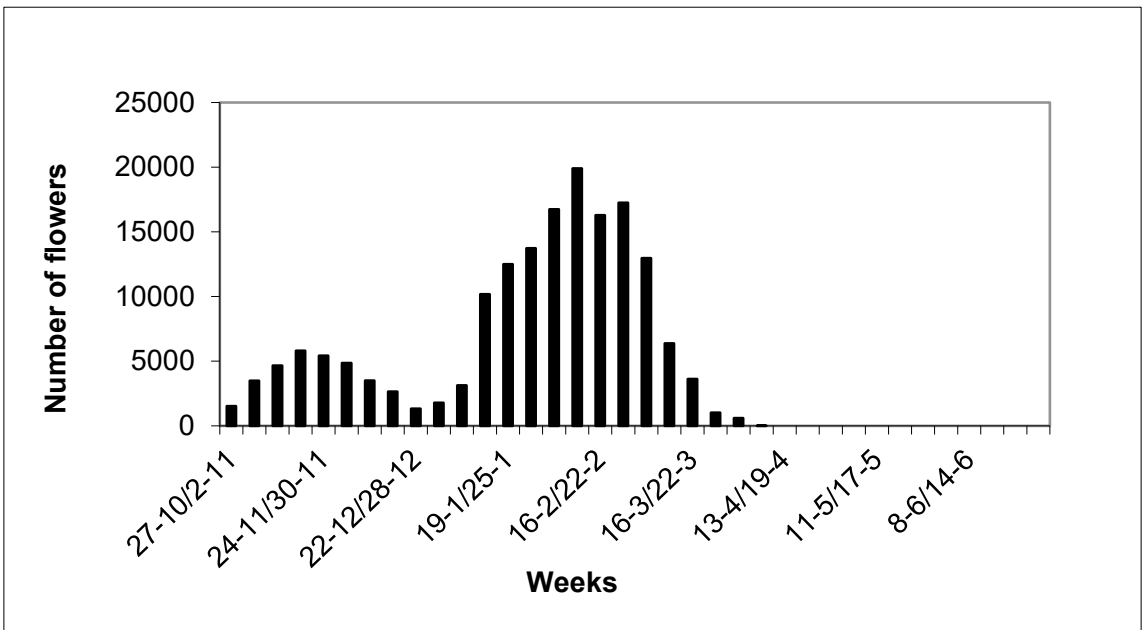

**Figure 5:** *Ulex parviflorus* flowers. From 27 October to 3 July of the second year.

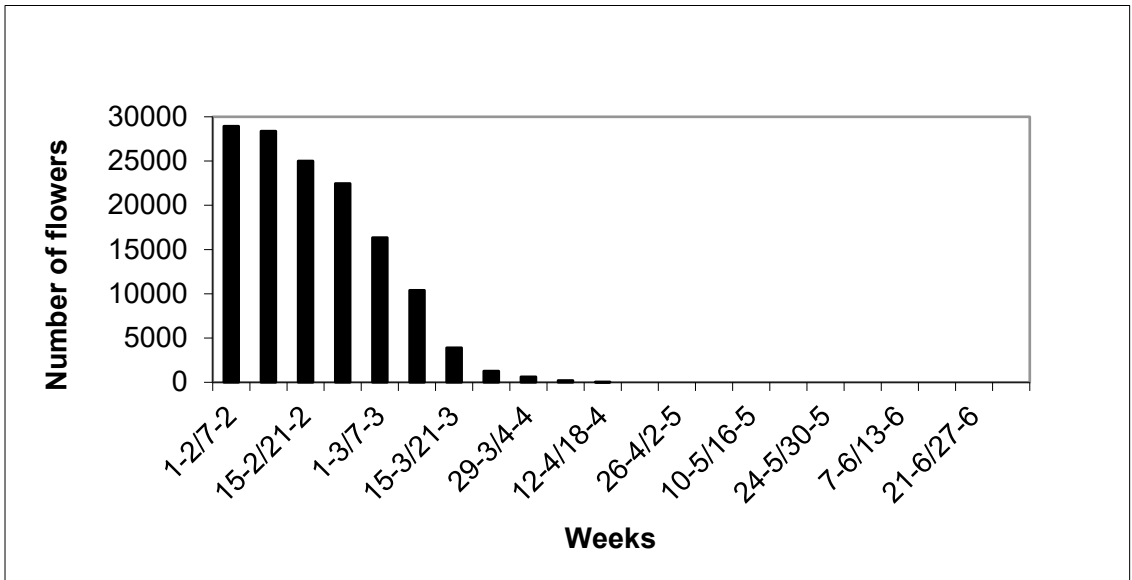

**Figure 6:** *Ulex parviflorus* flowers. From 1 February to 4 July of the third year.

### 3.3. Study of the evolution in *Ulex parviflorus* from flower to fruit.

Five specimens of *Ulex parviflorus* were selected at different stages of their development and a study was carried out on the evolution of 5 stages of development from flower to fruit, which are the following:

State 1: Flower bud.
State 2: Not open flower.
State 3: Open flower.
State 4: Dry corolla. The formation of the fruit begins.
State 5: Fruit formed.

The counting of the different states was realized in branches oriented to the north, south, east and west; with a size of 20 cm, with weekly frequency, from November of the first year to June of the second year.
In the study carried out on the evolution of the species *Ulex parviflorus* from the flower phase (stage 1) to the formation of a fruit (stage 5), it needs at least 6 weeks to complete its cycle.

In Table 2 we can observe the evolution of the chosen Ulex specimens, from the creation of the flowers to the ripening of the fruits. As we can see, the table includes, for each chosen Ulex specimen: (1) the number of each Ulex chosen, (2) the number of flowers that formed each Ulex, (3) the number of fruits formed by the flowers, (4) the percentage of fruits formed (number of fruits/number of flowers*100), (5) the number of ripened fruits and (6) the percentage of ripened fruits (number of ripened fruits/number of fruits*100). Finally, we calculated the mean and standard deviation of the percentage of fruit formed (4) and the percentage of ripened fruits (6).

From this information, we can conclude that the plant originates a great number of flowers, but not all develop as fruit. This occurs on average by 72.86%. The fruits formed mature 85.76%.

**TABLE 2**
**Evolution of the species of flower to fruit**

| Ulex | Number of flowers formed | Number of fruits formed | Percentage of fruits formed (%) | Number of ripened fruits | Percentage of ripened fruits (%) |
|---|---|---|---|---|---|
| 3 | 473 | 391 | 82,66 | 378 | 96,67 |
| 5 | 846 | 810 | 95,74 | 721 | 89,01 |
| 9 | 544 | 269 | 49,44 | 212 | 78,81 |
| 11 | 2074 | 944 | 45,51 | 870 | 92,16 |
| 18 | 772 | 686 | 88,86 | 495 | 72,15 |
|  |  |  | $\bar{X} = 72,86 \pm DT = 20,83$ |  | $\bar{X} = 85,76 \pm DT = 8,89$ |

## 4. POLLINATION

Pollination of *Ulex parviflorus* takes place by an external agent (the bees) that transport pollen.

### 4.1. Density of Bees

There are different methods to evaluate the density and activity of pollinators (bees) in relation to flowers. At first, the marking and retrieval technique for estimating the density of an animal population might seem appropriate (Wratten and Fry, 1982, pp. 37-43). A sample of the population was taken, marked and allowed to mix again with the rest, hoping that the proportion obtained in a subsequent sampling, between marked and unmarked animals, would be equal to the proportion of animals marked in the population. The more complex methods do not require a static population and allow the death of the animals during the management, as the method developed by Jolly (1956) is used for insect populations, but in the case of bees a problem arises. In order to mark the bees, the same markers used by the beekeepers to mark the queen were used, very colorful colors (red, blue, yellow ...) and which do not harm bees, but which cause disorientation of the bees, making it difficult for them to locate new sources of flowers and increasing slightly attacks by predators.

The previous study is very difficult to carry out, so a more accessible study has been made in our field observations (Dafni, 1992, pp. 165-225). The method used was to evaluate the relationship between bees' activity and their density with respect to the number of flowers. For this purpose, four specimens of *Ulex parviflorus* (Ulex 9, 11, 19 and 21) were randomly selected from the research plot.

The following counts were made:

a) Number of visitors per plant.
b) Number of flowers visited by bee.
c) Number of flowers available per plant.

The bees visit the flowers in the warmer hours of the day (11:00 a.m. to 4:00 p.m.) (Castañeda-Vildozola et al., 1999). The highest efficiency of pollination is usually between 10:00 am and 4:00 p.m. (Richards, 1987). The observation period was 15 minutes, during the hottest hours of the day (11 to 14 hours). The periodicity was weekly just after the period of maximum flowering (Dafni, 1992). The study was conducted during 9 weeks (February and March of the third year). The pollinators observed in the plot correspond to two species: *Apis mellifera* L. or common bee, is in charge of collecting the pollen of *Ulex parviflorus* in the experimental plot. Some specimens of *Bombus terrestris* L. have also been observed. The study of the period of flowering of this species from October to July, shows that the period of maximum flowering corresponds to February and March (Figures 7 and 8).

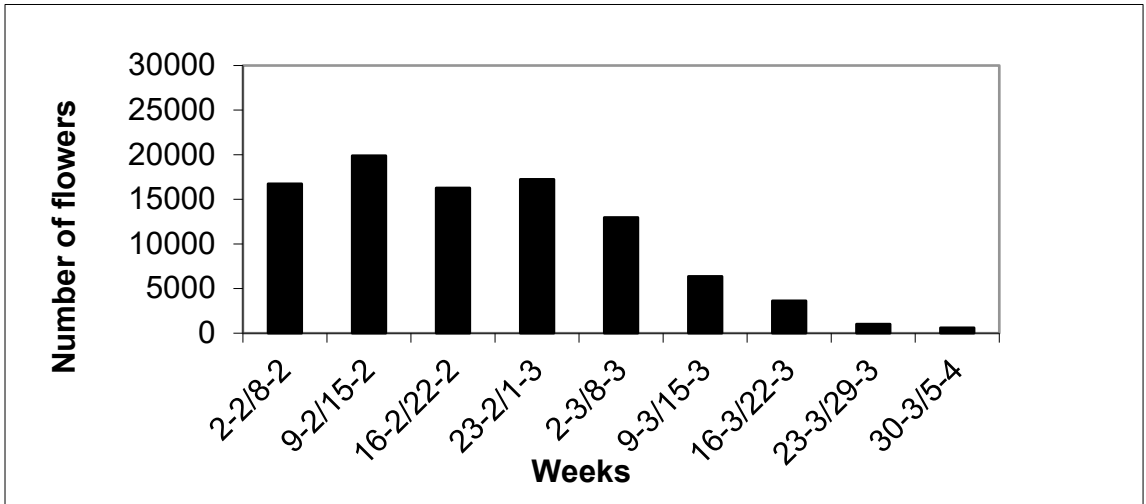

**Figure 7: Flowers of *Ulex parviflorus*. From 2-February to 3-April of second year.**

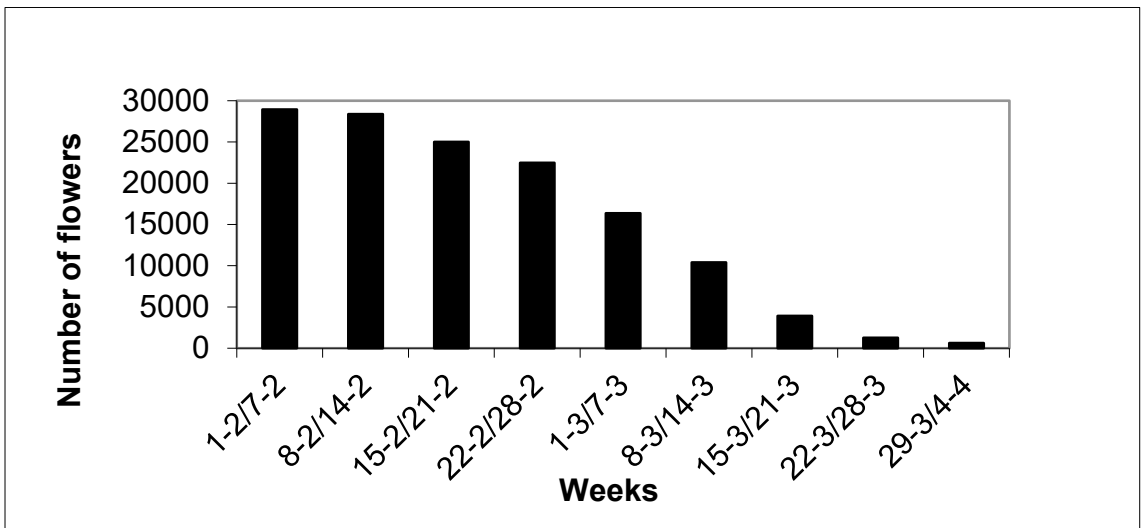

**Figure 8: Number of flowers of *Ulex parviflorus*. From 1-February to 4-April of third year.**

The meteorological factors studied in the plot during the experiment period were the mean temperatures and the average relative humidity of the air (Figures 9 and 10).

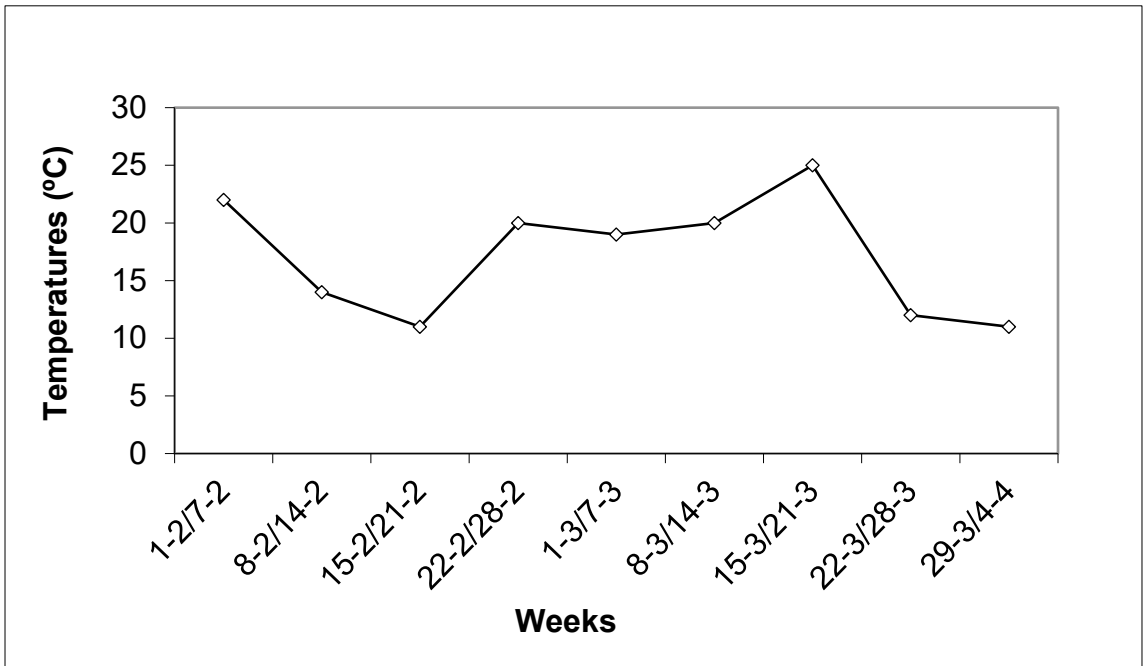

**Figure 9: Graph of average temperatures (11 to 14 hours) from 1-February to 4-April of third year.**

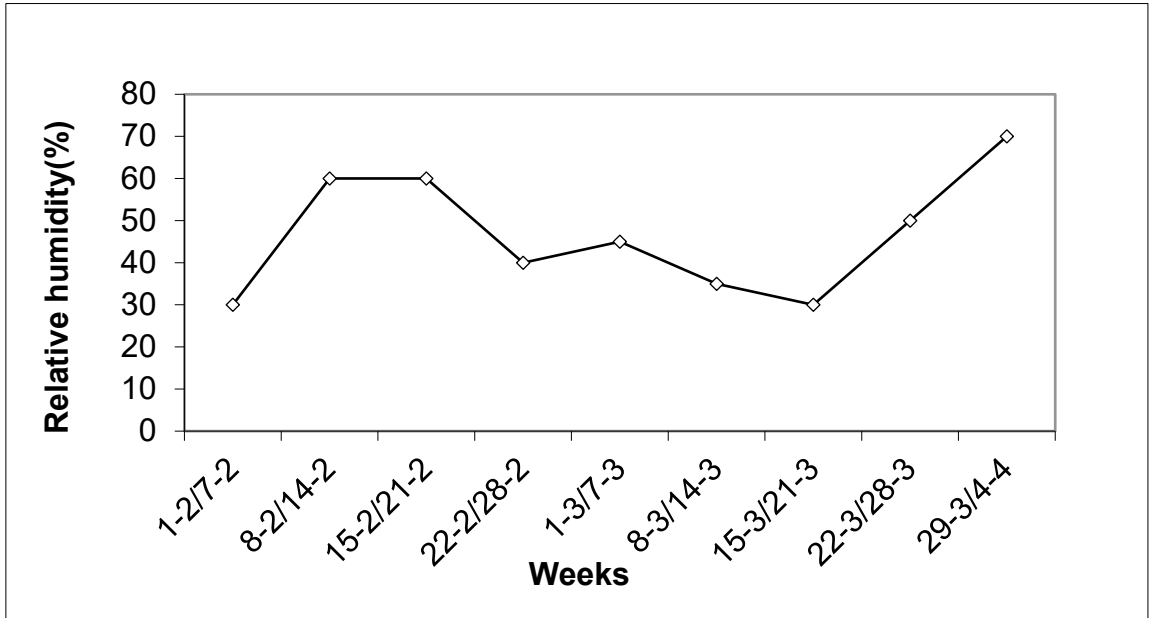

**Figure 10: Graph of RH (11 to 14 hours) from 1-February to 4-April of third year.**

The graphs below represent the number of flowers available in the four specimens of *Ulex parviflorus* studied and the number of bees/flowers visited (Figures 11 and 12).

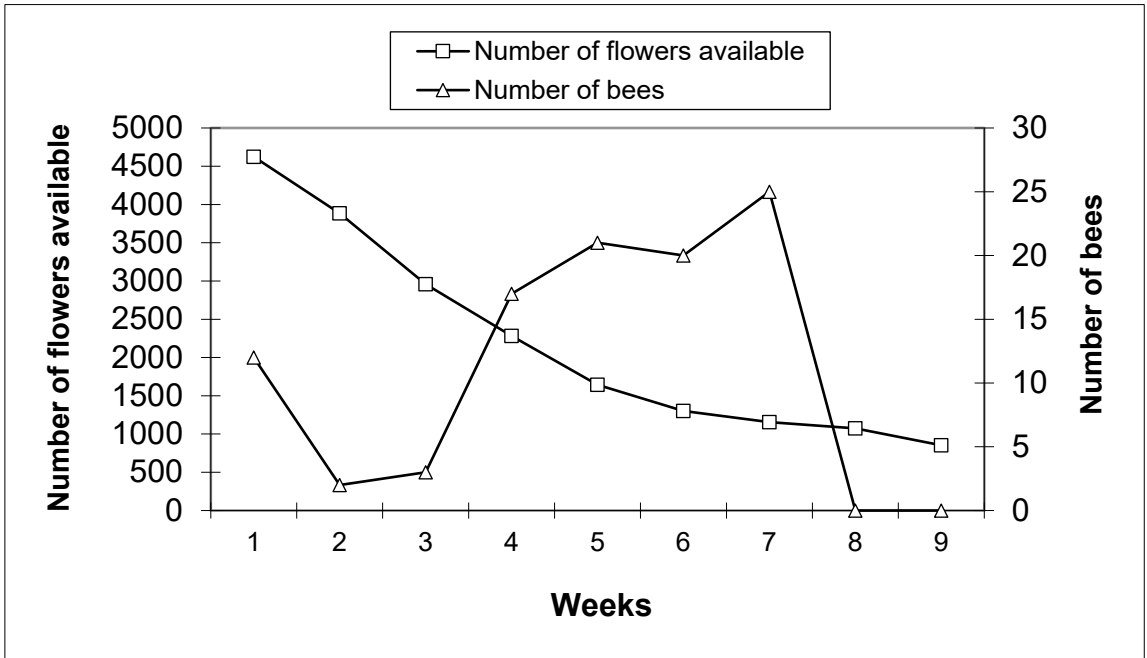

**Figure 11: Action of the bees from 1-February to 4-April of third year.**

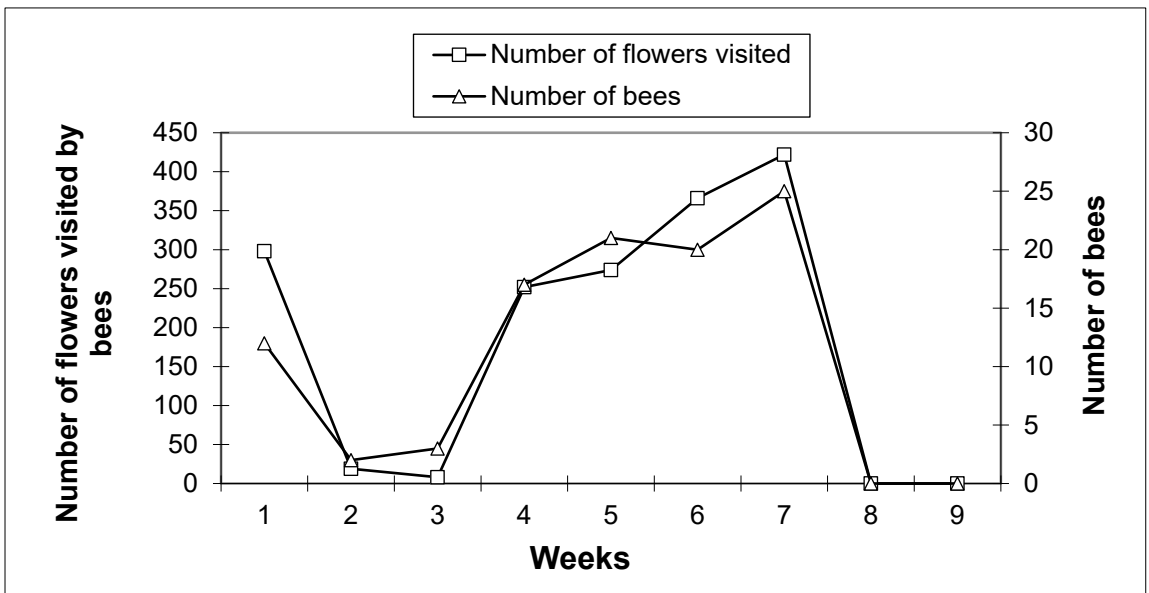

**Figure 12: Action of the bees from 1-February to 4-April of third year.**

Of the flowers present in plants, bees visit about 15% of these. The study carried out in the plot on the number of flowers of the plant, number of visitors per plant and number of flowers visited is reflected in the following results (Table 3):

## TABLE 3
### Study of the relation of the bees with the plot

| Observations (Weeks) | $\overline{X}$ visits/plant | $\overline{X}$ flowers visited | $\overline{X}$ flowers available |
|---|---|---|---|
| 1 | 3 | 24'8 | 1156 |
| 2 | 0'5 (*) | 9'5 | 971 |
| 3 | 0'75 (*) | 2'6 | 739'5 |
| 4 | 4'25 | 14'82 | 570'5 |
| 5 | 5'25 | 13'04 | 411 |
| 6 | 5 | 18'3 | 325'5 |
| 7 | 6'25 | 16'88 | 288'75 |
| 8 | 0 (*) | 0 | 268'25 |
| 9 | 0(?) | 0 | 213'25 |

(*)It was very windy
(?)Cloudy day

Observations have been made in consecutive weeks.
The visit rate index is equal to the total number of visits during the observation period/number of flowers available during that period (Zimmerman[a], 1980)
The effectiveness of the visit rate per flower and per unit of time is (Andersson, 1988):

$$V = (A \times N)/C$$

A = Number of visitors per plant per unit of time.
N = Number of flowers visited per unit of time.
C = Number of flowers per plant.

The index of attraction is equal to the number of visitors/flowers available/unit of time (Pleasants, 1980).
The efficiency of pollination is equal to the number of flowers visited per unit of time (Richards, 1987).
The indices studied show the following data (Table 4):

## TABLE 4
### Table of índices

| Observations (weeks) | Visiting rate | V | Attraction index | Efficiency of pollination |
|---|---|---|---|---|
| 1 | 0'064 | 0,773 | 0'0025 | 19'86 |
| 2 | 0'004 | 0'009 | 0'0005 | 1'266 |
| 3 | 0'002 | 0'006 | 0'0010 | 0'400 |
| 4 | 0'110 | 1'877 | 0'0074 | 16'80 |
| 5 | 0'166 | 3'5 | 0'0127 | 18'26 |
| 6 | 0'281 | 5'622 | 0'0153 | 24'40 |
| 7 | 0'365 | 9'134 | 0'0216 | 28'13 |
| 8 | 0 | 0 | 0 | 0 |
| 9 | 0 | 0 | 0 | 0 |

## 5. FRUCTIFICATION

### 5.1. Characteristics of the fruit

The fruits are dry, in legume, with one or two seeds. The size ranges from 8.5-9 x 4-4.5 mm, as long as the calyx or somewhat larger, oval-oblong (Castroviejo, 1999) (Figure 13).

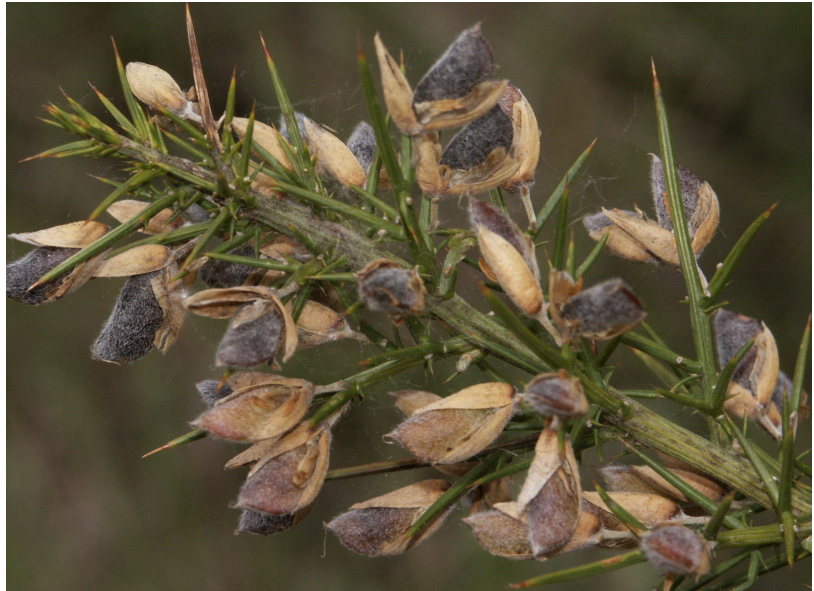

**Figure 13: Fruits of *Ulex parviflorus*.**

The seed is hard, smooth and usually of intense yellow color, ranging from green to dark brown; has a lenticular shape, without appendages except a protrusion called an aryl, through which the fruit is inserted (Figure 14).

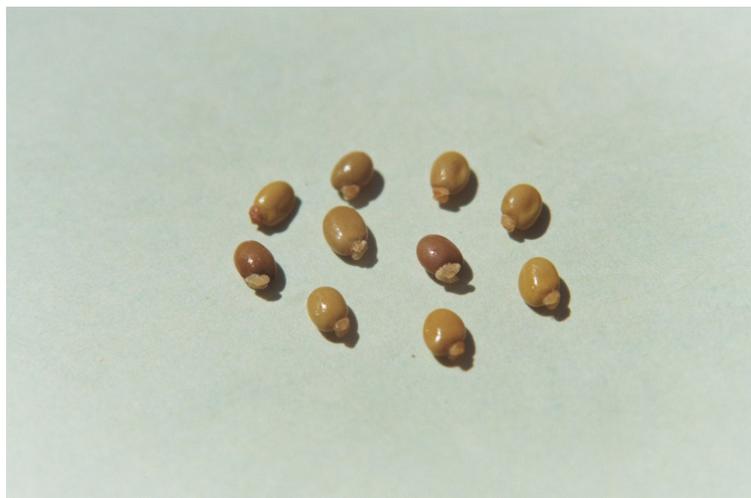

**Figure 14: Seeds of *Ulex parviflorus*.**

The aryl is part of the diet of ants, which is a means of secondary dispersion of the seeds. The seeds have a size ranging from 2.5-4 mm in diameter and 1.6-3 mm in diameter with a weight ranging from 0.0045-0.006 grams (Baeza et al., 1991).

## 5.2. Study of fruiting

Fieldwork has been carried out for three consecutive years, distributed as follows, the first and second year of sampling, from October to July of the first year. From October to July of the second year, fruit counts were carried out throughout the fruiting period and the third year, from February to July, the evolution from the beginning of maximum fruiting was studied until its disappearance. The study carried out on the fruiting period of this species from October to July is based on the fact that from November, December, the fruits begin to form, reaching the highest number of fruits in the plant in March, April. From this period the expulsion of the seeds from the fruits begins. The primary dispersion period runs from late March to June, presenting a period of maximum dispersion in June (Figures 15, 16 and 17).

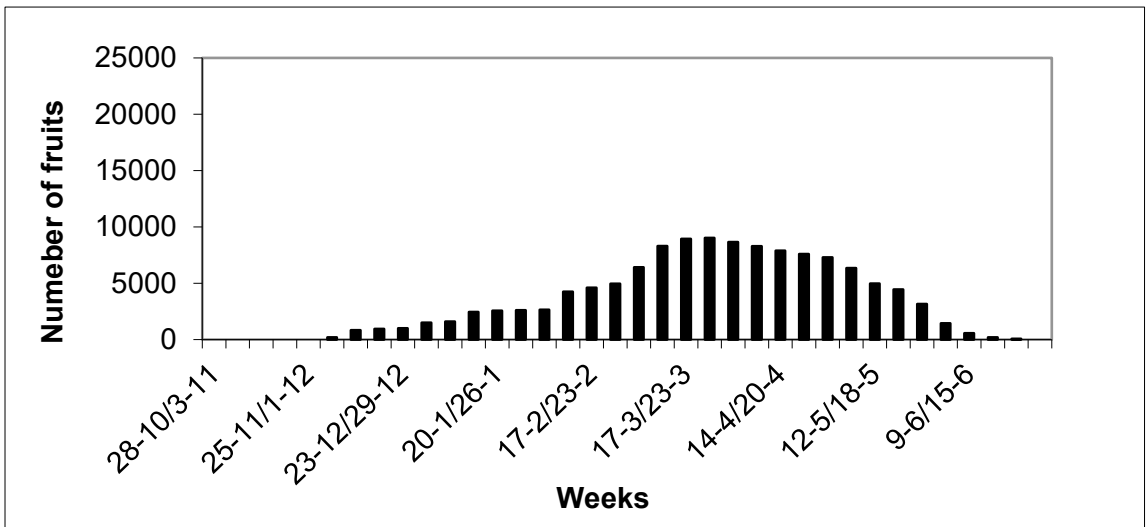

**Figure 15: Fruits of *Ulex parviflorus*. From 28 October to 6 July of first year.**

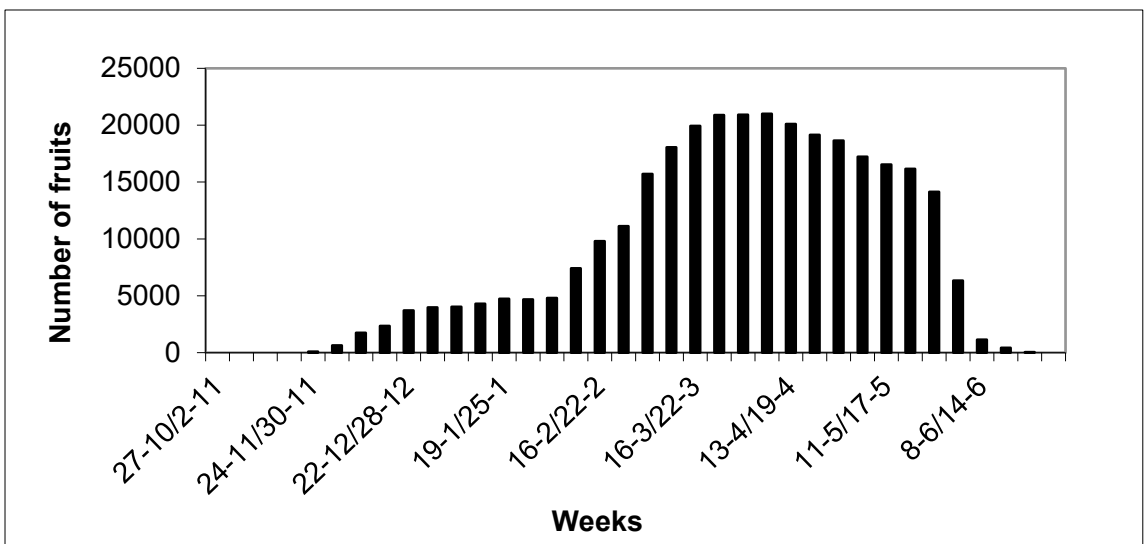

**Figure 16: Fruits of *Ulex parviflorus*. From 27 October to 5 July of second year.**

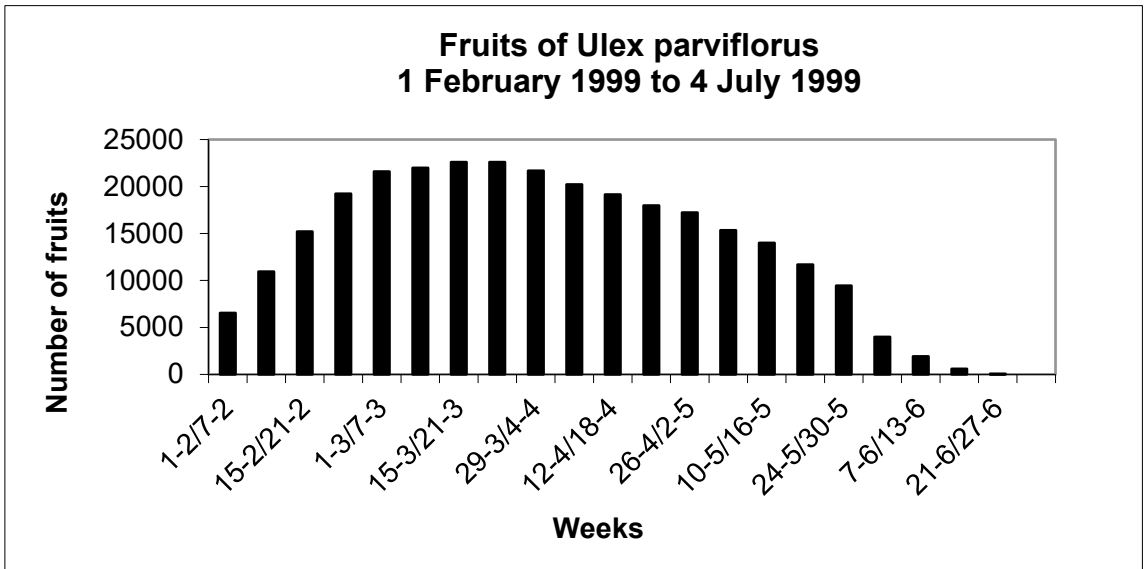

**Figure 17:** Fruits of *Ulex parviflorus*. From 1 February to 4 July of third year.

## 6. SEED DISPERSAL

The species *Ulex parviflorus* is an obligate germinating species, which does not regenerate by regrowth and presents a large amount of accumulated necromass at the base of the shrub. Mean distribution of seeds in soil (with respect to distance and orientation) (Figures 18 and 19).

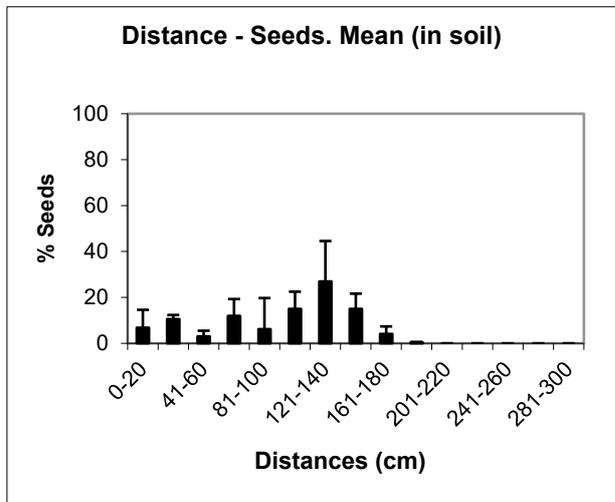
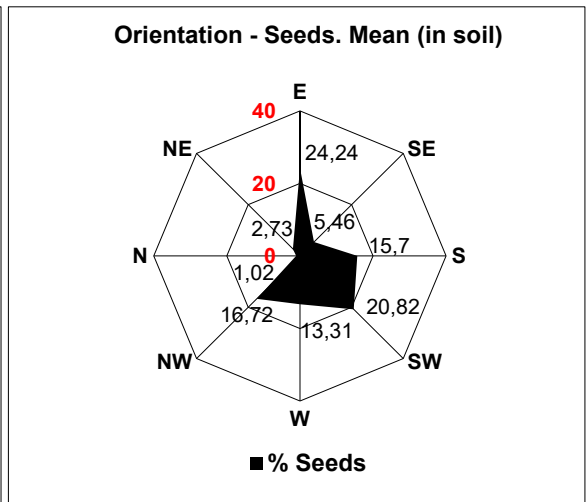

**Figure 18**                                                **Figure 19**

The obtained data show a greater amount of fruit distributed in the middle of the plant and a greater dispersion of the seeds in the soil between 140-150 cm., also a lack of coincidence is observed in the orientations of the fruits in the plant and the situation of the seeds in the soil, influenced by the speed and direction of the wind.

Preliminary evidence indicates that seed production in the 5- to 7-year-old formations is much higher than in the more mature ones. This greater production will contribute to increase the seed banks, stabilizing when the production reaches the state of maturity. Therefore, control treatments of gorse at intermediate ages, before maximum seed production, could contribute to reduce the seed bank and the regenerative capacity of the species.

After dispersion of the seeds, the plant enters a period of rest, until the middle of autumn when the flowering begins again. The fruiting period of this species reflects that the period of primary dispersion runs from the end of March to June, presenting a period of maximum dispersion in June. The seeds of *Ulex Parviflorus* are violently expelled from their fruits (explosive dehiscence) that then fall to the ground by gravity. If only this dispersion existed, all seeds should be located close to the parent plant, but this does not occur with the possibility of secondary dispersion.

## 7. ROLE OF ANTS IN THE DISPERSAL OF THE SEEDS

When studying the dispersion of the seeds of *Ulex parviflorus* it is verified that the structure of the fruits is intimately related to its mode of dissemination, presenting an explosive dehiscence (autochory). If the seeds of *Ulex Parviflorus* are violently expelled from their fruits, they must accumulate in the soil, around the mother plant, but from direct observation this does not occur. On the contrary, the number of seeds has been reduced considerably. Where are the seeds? Abad et al. (1996) say that obligate sprouting shrubs, among which this species is found, have persistent seed banks in the soil. Who creates these seed banks?

The seeds of *Ulex parviflorus* have an appendix. Sernander (1906) was the first to analyze this appendix of some seeds called an aryl or elaiosome, proving that it has a biochemistry, size and shape different from the rest of the seed. An elaiosome is a special external adaptation of seeds to improve ant dispersal (Brew, O'Dowd and Rae, 1989; Ohkawara and Higashi, 1994). This external structure contains nutritive reserves such as lipids, proteins, starch, sugars and vitamins and is present in some diaspores of zoochore plants. Animals that intervene in its dissemination, mainly ants, seek these diaspores to use their aril as food, while the rest of the seed is abandoned without having suffered a noticeable alteration. Granovarian ants use the aryl to pick up the seed and carry it to the anthill, which will feed other worker ants or larvae, while the seeds will remain intact, usually in the anthill (Bennett and Krebs, 1987; Beattie, 1985).

Ants help the germination of the seeds, since they add to the soil surrounding the anthills nutrient content and different physical and chemical properties (Beattie and Culver, 1982). They disperse seeds and decrease intra-specific competition as a result of the occupation of space. (Gorb and Gorb, 1999). Myrmecochory is a mutualism (Handel and Beattie, 1990) in which ants use the nutritive value of elaiosomes as food and disperse plant seeds at a distance from the source plant (Andersen, 1988; Retana et al., 2004). It occurs almost all over the world appearing in sclerophyllous forests (Berg, 1975; Westoby et al., 1982). Botanists who have analyzed the mycorrhiza have concluded that the scarcity of nutrients, particularly phosphorus and potassium, is the key to uneven geographical distribution (Westoby et al., 1982). The existence of a complicated structure with an apparent function that is none other than the attraction of the ants implies that the myrmecochory accumulates a substantial selective advantage. Botanists have dealt with this advantage and have found five possibilities:

1) Avoid inter-species competition
2) Avoid fire (Berg, 1975). Myrmecochory is common in pyrophytic plants. It is possible that the ants carry the seeds to the anthills, to protect them during the frequent fires of that habitat.
3) Avoid parental competition.
4) Prevent seed predation.
5) Micro-sites with higher nutrients (Hölldober and Wilson, 1990)

The questions we raise, the hypotheses and the predictions are as follows:

1) Are the seeds of *Ulex parviflorus* secondarily dispersed by ants in an active way? If so, the ants:
a) They would be attracted to aryl and therefore prefer aryl seeds to those that do not.
b) They would look for the seeds more intensely where there would be more possibilities of success and therefore the activity of collection by the ants will be related especially with the pattern of the primary dispersion.

2) Have positive consequences for *Ulex parviflorus* this interaction with ants? If so:

a) The seeds of *Ulex parviflorus* without aryl should germinate more and/or faster than with aryl.
b) Seeds of *Ulex parviflorus* without aryl should be found in the anthill dump.

3) How important are ants to *Ulex parviflorus*?

We will ask the following questions:

### 7.1.1. Do ants prefer aryl seeds?
To verify if the ants had any preference in the choice of seeds, the following experiment was performed: Five specimens of *Ulex parviflorus* were randomly selected in the experimental plot (Ulex 1, 5, 7, 8, 15). Later, 4 closed Petri plates were placed in each plant, with two opposing holes of 1 cm, covered with weights to avoid being carried away by the wind. In the plates were placed: 1 plate with 6 seeds with aryl (6 CA), 1 plate with 6 seeds without aryl (6 SA) and 1 plate with 3 seeds with aryl and 3 seeds without aryl (3 CA 3 SA). Varying the orientation (N, S, E, W, SE, SW, NE, NW) and distances (between 60 and 150 cm.). The plates were observed weekly and the numbers of remaining seeds were recorded all had gone (seeds were not replenished). The experiment was repeated from March to June, although only data from the first 5 weeks were used, since the ants seemed to learn the location of the plaques from that date. The variable used was the number of "surviving" seeds per week which was analyzed by the Kaplan-Meier analysis.

### 7.1.2. Are special patterns of primary seed dispersal and collection by ants related?
In order to be able to say that the seeds of *Ulex parviflorus* are dispersed by the ants it must be demonstrated that they preferentially look for the seeds of *Ulex parviflorus* in the places and times where they accumulate most. The questions raised were: Where do seeds accumulate? When? Where do ants look for seeds?

A) In order to know when the primary dispersion occurs and of which environmental factors it is necessary to count the number of fruits closed in the plants of the plot. Since each fruit of *Ulex parviflorus* normally has one seed, except in a few exceptions which may have two or even more, a reliable measure of the dispersion is counting the fruits. The study was carried out during three years, during periods of maximum fructification (March to July). During the period from the end of March to June of the third year, the variables that may affect the release of the seeds from their pods were hypothesized to be temperature and relative humidity, which were also taken into account, given the hygroscopic nature of the force of opening of the fruits. The comparison was made based on the number of fruits closed and the hours per week that the plants have been above a certain temperature (10 ºC, 15 ºC, 20 ºC, 25 ºC and 30 ºC), and the number of hours per week which have been below a certain relative humidity (30%, 40%, 50%, 60% and 70%), so that we could obtain the best predictor of the opening of the fruit.

B) For the study of the primary dispersion pattern of seeds in the soil, three specimens of *Ulex parviflorus* (Ulex X, 9, 20) were randomly selected, two specimens of the experimental plot and another (X) covered the soil with a 3m x 3m cloth, with a 0.5 mm diameter web, to collect the dispersed seeds. To avoid modification of seed dispersal by ants, a repellent was used. The distance from the seeds on the ground to the trunk of the plant was measured and also the orientation with respect to the plant. The periodicity of measurement was weekly from the period of maximum fructification (April to June of the second year).

C) Do ants search for seeds by following the pattern of distances and orientation with which seeds are dispersed, or is the ants' behavior random? Three specimens of *Ulex parviflorus* (Ulex 3, 9, 20) of the thesis plot were selected. Five mature seeds (with aryls) were placed in each Petri dish and with two holes of 1 cm in diameter at each end of their bases (covered with weights to avoid the force of the wind and allow the passage of the ants) at different distances and orientations. The plates were placed at 30, 90, 150 and 210 cm and in the orientation: E, W, N and S. The number of seeds remaining on the plates was counted each week. Once counted, seeds were added so that there were 5 seeds on the plate, the distance was maintained and the orientation of the plates was rotated clockwise relative to the plant trunk (SE, SW, NE, NW) , to prevent the ants from learning their location. The experiment lasted 8 weeks (March to June of the third year). The variable analyzed was the percentage of predation and was analyzed by an analysis of the univariate variance, with the factors being the distance and the orientation. Since the data did not meet the normality requirement, even after applying a transformation, a nonparametric variance analysis (Kruskal-Wallis test) was performed separately for each factor (distance and orientation).

### 7.1.3. Is the germination rate the same for seeds with or without aryl?

Seeds of *Ulex parviflorus* were collected from the experimental plot (with and without aryl from the soil) and collected by the ants (with and without aryl): These seeds were planted in Petri dishes 9 cm in diameter, on filter paper and 5 ml. of distilled water (25 seeds per dish) and left to germinate in a chamber with cycles of 12 light/dark hours at 20º/10º C respectively. This experiment was carried out from January 4 of the third year (sowing day) until April 25 of the same year. The data analysed using ANOVA to test the effect of aryl, using the germination rate as the dependent variable. Given that in order to perform these analyzes it is necessary that the dependent variable used follow a normal distribution with homogeneous variance, i.e., that they fulfill the normality and homogeneity assumptions of the required variance in ANOVA. We proceeded to perform the Kolmogorov-Smirnov normality test and the Levene contrast test on the equality of error variances. From such tests it appears that the germination rate, fulfills the assumptions of normality and homogeneity of the variance and can apply parametric tests.

### 7.1.4. Where do they go for those seeds collected by the ants?

To demonstrate where the seeds of *Ulex parviflorus* are found and to infer the role of ants in this process, the following study was carried out in May of the third year. First the ants of the zone were located. Afterwards, the contents of all the ants' garbage dumps were studied. But only the interior of a single anthill was studied to search for the presence of seeds in the store (the reason for analyzing a single anthill was to avoid the modification of the study area to the maximum).

### 7.1.5. What is the importance of ants on *Ulex parviflorus*?

The ants play an important role in the study of seeds, so the work is completed by carrying out a study of these ants, their density in the plot and possible ant mounds in the area, so that one can get an idea of the impact in absolute terms of the ants on *Ulex parviflorus*.

a) **Study of ants**. The different types of ants observed in the plot were collected in hermit seals (type Ependorf) with alcohol at 60º and sent to the Department of Animal Biology and Ecology of the University of Granada (Spain), for identification.
b) **Density of ants**. To calculate the density of ants (workers) in the soil, the passive method of installing fall traps was used (Levieux, 1969). The fall traps consisted of containers 9 cm in diameter and 1,5 cm in height and with a capacity of 60 milliliters and a time of evaporation of approximately 48 hours, inside which only water was introduced to avoid attraction or repulsion and a few drops of detergent to remove surface tension. Ten drop traps were placed in two parallel rows, buried to the edge and separated from each other by about 3 meters. The count of the captured ants was done every 24 hours of the stay in the field. The experiment was repeated for 5 weeks between the last week of May and the first of July of the third year.
c) **Presence of ants in the area**. The anthills were counted to identify the ants' species and the anthills' density.

### 7.1.6. Do ants prefer aryl seeds?

The results obtained on the preference of the ants with respect to the seeds with or without aryl are the following (Table 5):

**TABLE 5**

|  | Survival time $\bar{X} \pm DT$ | 95% Confidence interval |
|---|---|---|
| CA (with aryl) | 9,80 ± 3,42 | (6,44 – 13,16) |
| SA (without aryl) | 25,20 ± 10,48 | (14,93 – 35,47) |
| CA (CA+SA) | 16,80 ± 8,40 | (8,57 – 25,03) |
| SA (CA+SA) | 25,20 ± 8,40 | (16,97 – 33,43) |

Log Rank Statistic for CA vs. SA is as follows: Log-Rank $\aleph^2 = 5,33$; p = 0,0210

Log Rank Statistic for CA (CA+SA) vs. SA (CA+SA) is as follows: Log-Rank $\aleph^2 = 1,79$; p = 0,1811

The results obtained indicate an attractive effect of aryl (a higher preference of ants for seeds with aryl), but this effect is statistically significant only when the seeds are presented in separate traps, but not when presented in the same trap (Figures 20 and 21).

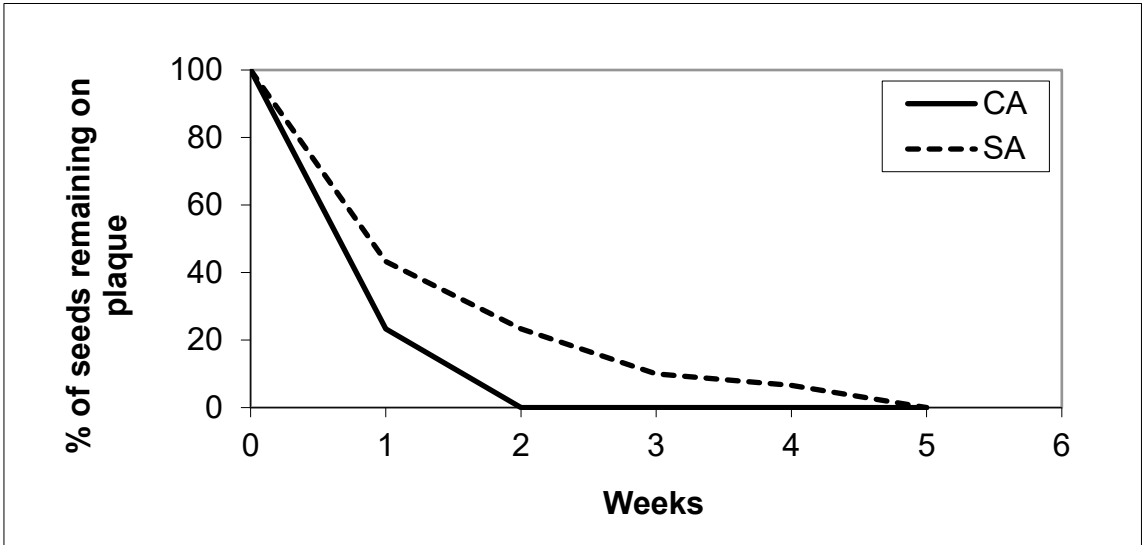
**Figure 20: Survival rate of seeds on plaque.**

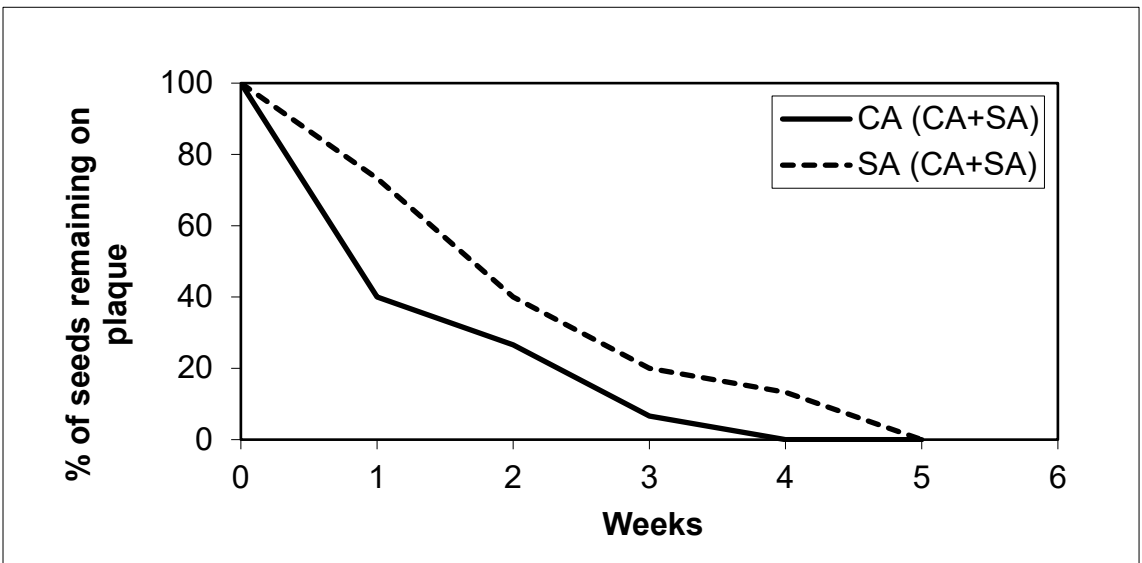
**Figure 21: Survival rate of seeds on plaque.**

**7.1.7. Are the special patterns of primary seed dispersal and ant harvesting related?**
1) The study carried out on the fruiting period of this species from March to July, shows that the period of primary dispersion runs from the end of March to June, presenting a period of maximum dispersion in June, this was consistent in the three years of study (Figures: 22, 23 and 24).

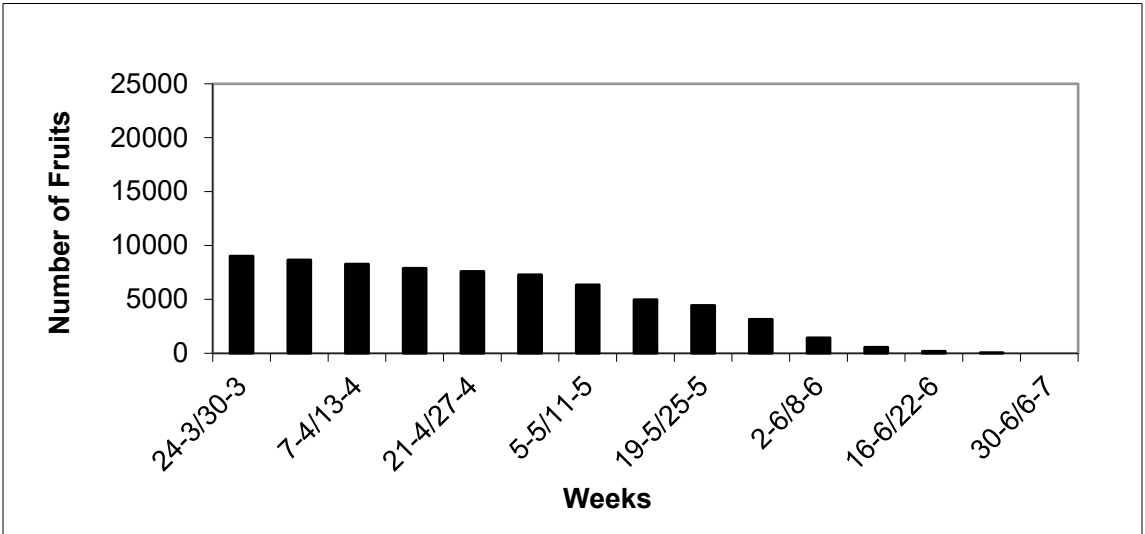

**Figure 22: Fruits of *Ulex parviflorus*. From March 24 to July 6 of the first year.**

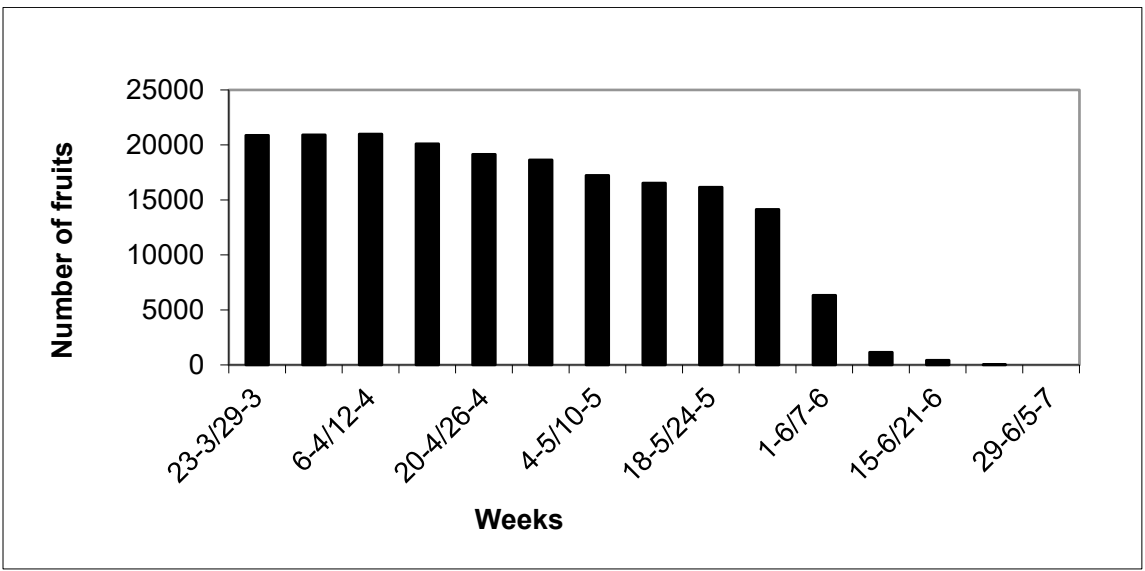

**Figure 23: Fruits of *Ulex parviflorus*. From March 23 to July 5 of the second year.**

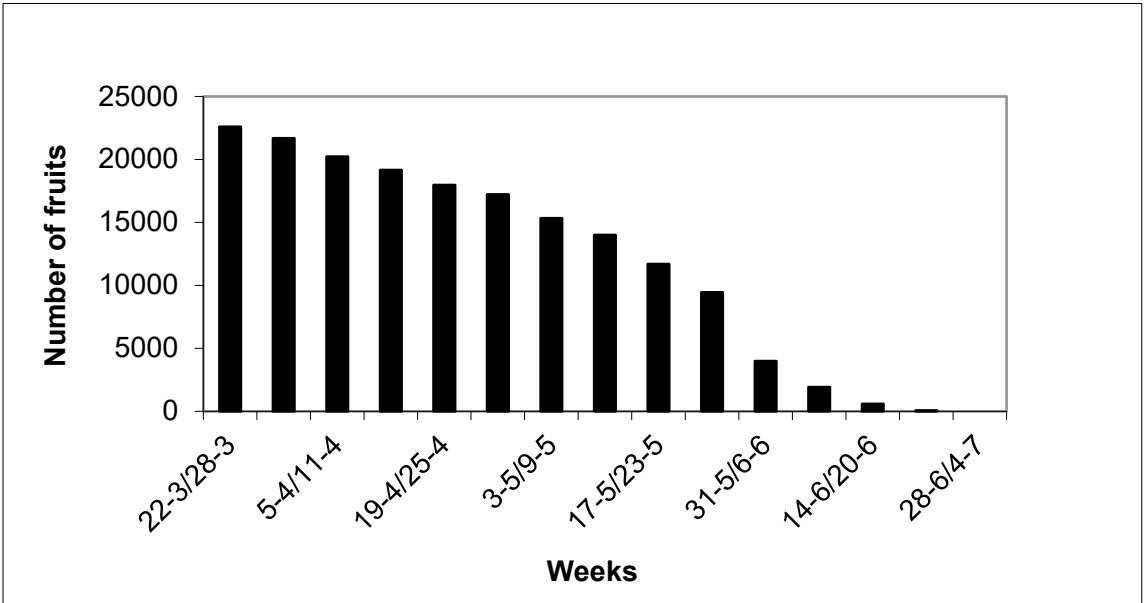

**Figure 24: Fruits of *Ulex parviflorus*. From March 22 to July 4 of the third year.**

When the data obtained are related to the environmental conditions, they indicate that the period of maximum dispersion starts and accelerates coinciding with moments of low relative humidity and an increase of the ambient temperature, as expected. The curves that best seem to predict this process are Temperature $\geq$ 15 ºC and Relative Humidity $\leq$ 70% (Figures 25 and 26). Figure 25 has been obtained by measuring weekly the number of hours and fruits that have been above the temperatures 10ºC, 15ºC, 20ºC, 25ºC and 30ºC. On the other hand, figure 26 has been obtained by measuring weekly the number of hours and fruits that have been below a relative humidity of 30%, 40%, 50%, 60% and 70%.

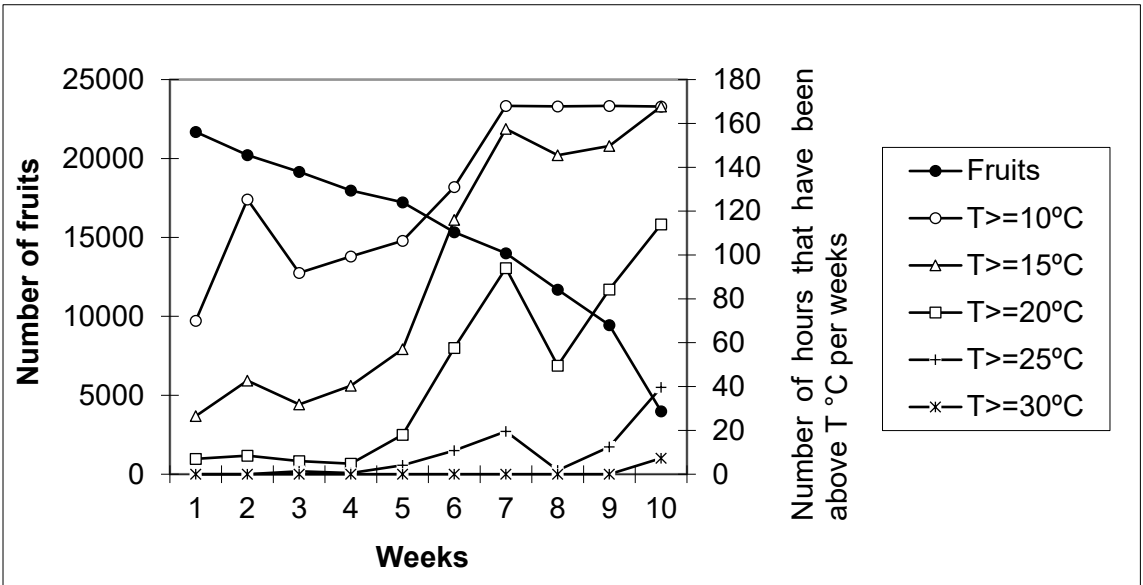

**Figure 25: Temperatures: March – June of the third year**

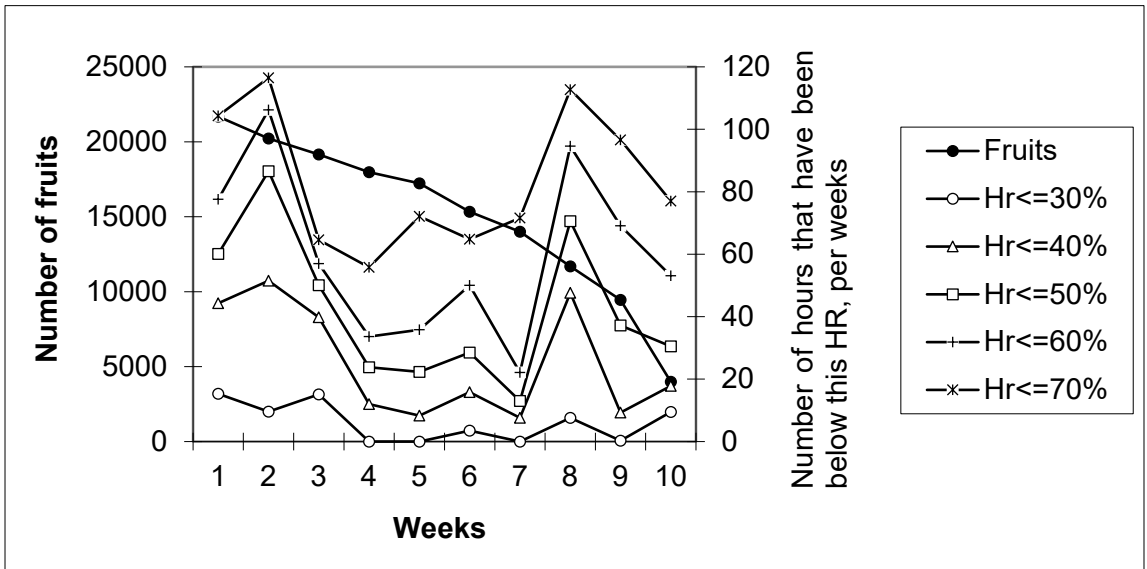

**Figure 26: Relative humidity: March – June of the third year**

2) The results of the study on the spatial distribution of fruits in height and orientation, as well as the dispersion of the seeds in the soil with respect to distance and orientation, reflect the following results (Figures: 27, 28, 29 and 30):

a) Mean (distribution of fruits in the plant by height and orientation)

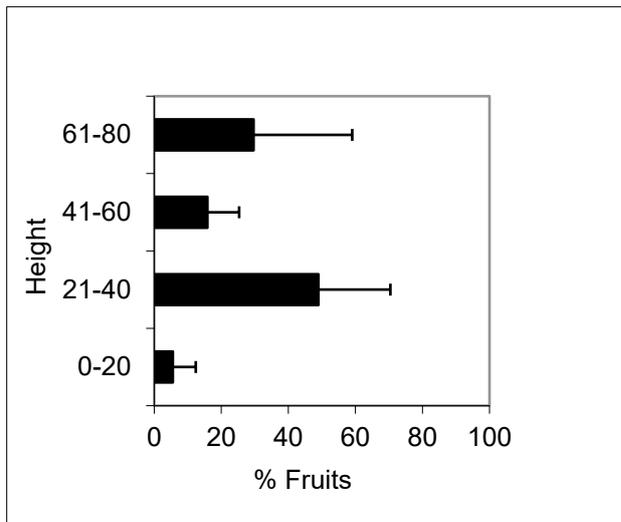 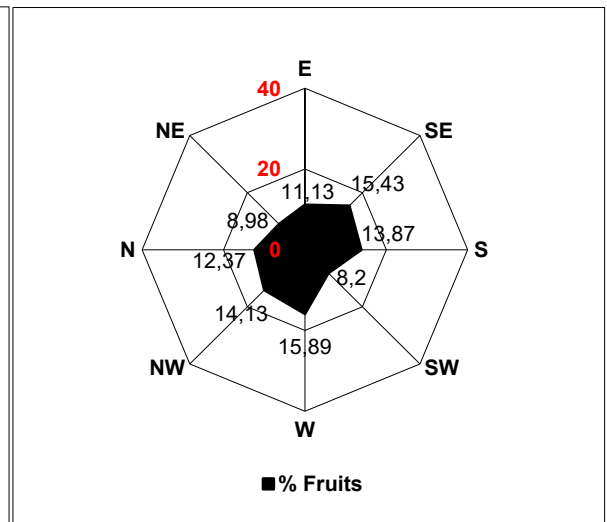

**Figure 27: Height - Fruits. Mean**     **Figure 28: Orientation - Fruits. Mean**

b) Mean (distribution of seeds in soil with respect to distance and orientation)

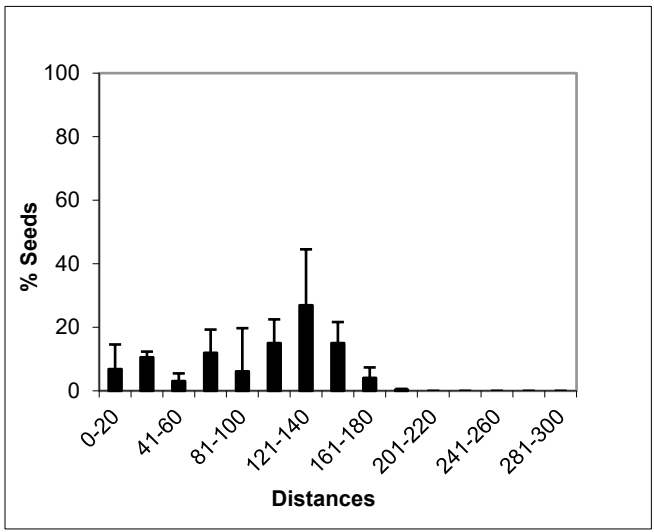
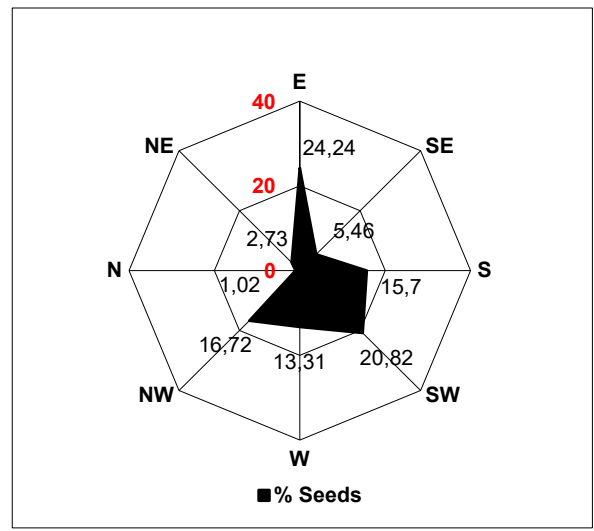

Figure 29: Distance - Seeds. Mean (in soil)  Figure 30: Orientation-Seeds. Mean (in soil)

The data obtained show a greater amount of fruit distributed in the middle of the plant and a greater dispersion of the seeds in the soil between 140-150 cm. Also a lack of coincidence is observed in the orientations of the fruits in the plant and the situation of the seeds in the soil, influenced by the speed and direction of the wind.

### 7.1.8. Do the ants follow the distance and orientation pattern of the primary seed dispersal, or is it random?

The study of the search for seeds by the ants at distances (30, 90, 150 and 210 cm) and in all orientations reflects the following results (Table 6):

**TABLE 6**

| TRDIST | | TRTOR | |
|---|---|---|---|
| Distance | $\bar{X} \pm DT$ | Orientation | $\bar{X} \pm DT$ |
| 30 cm | 67,10 ± 29,02 | N = NE | 73,80 ± 30,36 |
| 90 cm | 75,00 ± 31,01 | E = SE | 65,00 ± 33,39 |
| 150 cm | 71,30 ± 29,72 | S = SW | 66,70 ± 34,60 |
| 210 cm | 61,30 ± 37,45 | W = NW | 69,20 ± 30,31 |

The clustering variable: DISTANCE (Figure 31), gave the following result according to the Kruskal-Wallis test, Ĥ = 4,672; df = 3; P> 0.05.

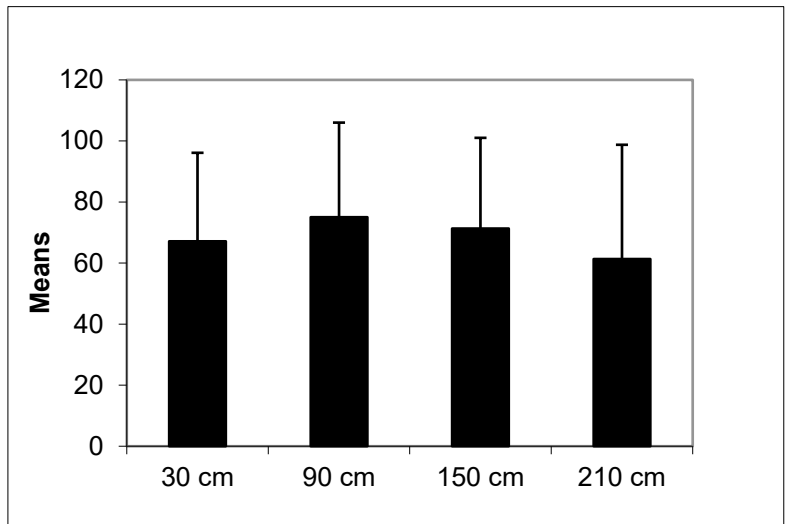

**Figure 31: Distances.**

The cluster variable: ORIENTATION (Figure 32), gave the following result according to the Kruskal-Wallis test, Ĥ = 2.056; df = 3; P> 0.05.

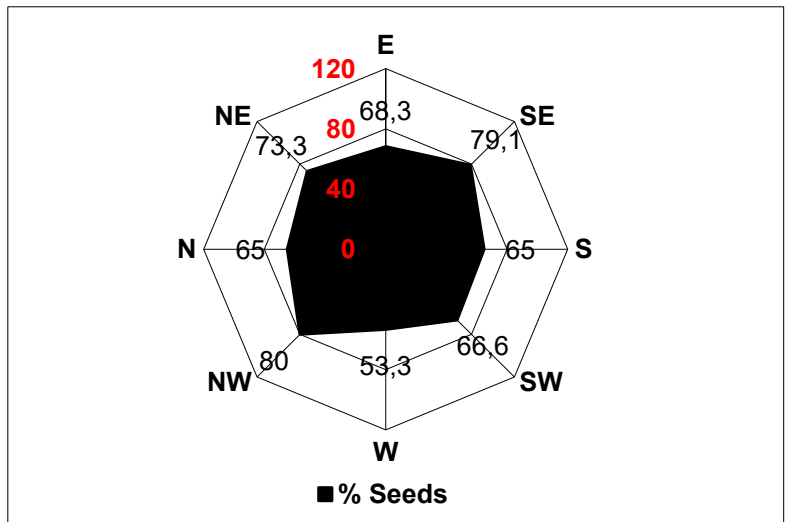

**Figure 32: Orientation.**

In the relationships between distance (30, 90, 150, 210) and orientation (E, SE, S, SW, W, NW, N, NE) no significant differences were found. From which it is inferred that the ants do not show a pattern of foraging, optimized for the search of the seeds of *Ulex Parviflorus*.

### 7.1.9. Germination of the seeds of *Ulex parviflorus*
The results obtained in the analysis of the effect of the aryl on the germination rate indicate a significant effect with a higher value for the seeds without aryl (SA) (Table 7).

**TABLE 7**

| Treatment | $\overline{X} \pm DT$ (Rate on germination) |
|---|---|
| CA (With aryl) | 19,60 ± 8,52 |
| SA (Without aryl) | 33,20 ± 9,80 |

The germination rate in the first month of germination of the seeds without aryl (SA) surpasses the seeds with aryl (AC) at 1.5%, from the first month and until the second 627 month the difference increases and then remains almost stable (Figure 33).

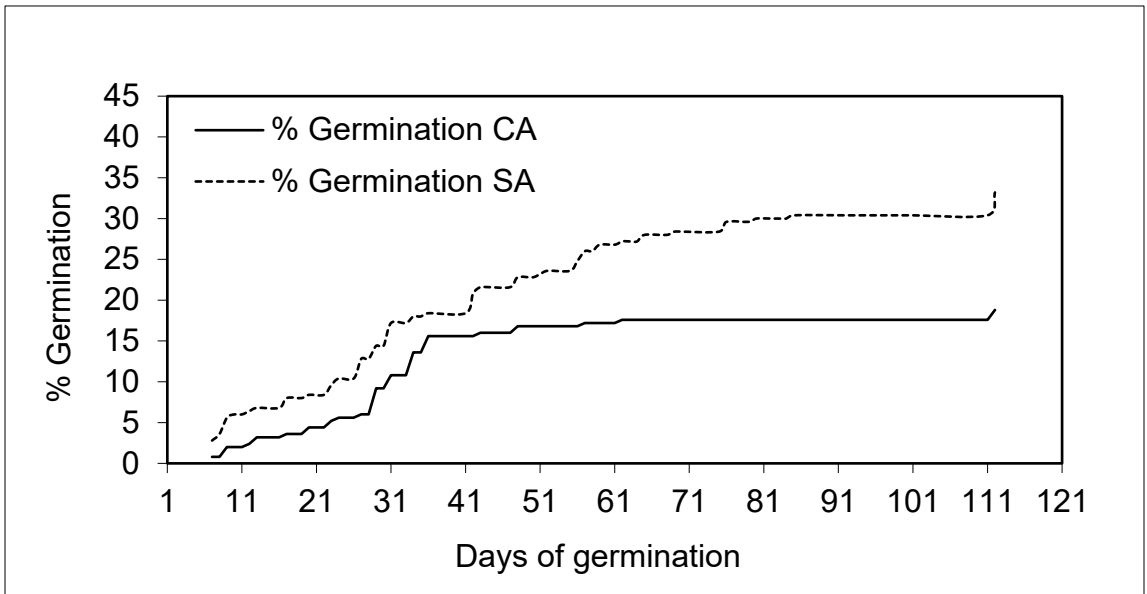

**Figure 33:** Germination of *Ulex parviflorus*

### 7.1.10. Where will the seeds of *Ulex parviflorus*, collected by the ants?

In the analysis carried out on 6 anthills of *Messor barbarus* (L.), it was observed that all the ants' external garbage dumps contained seeds of *Ulex parviflorus* ("with" and "without" aryl). Subsequently the study of an anthill was carried out and an accumulation of seeds at 2 to 4 cm of depth was found. These seeds were also "with" and "without" aryl (Table 8).

**TABLE 8**

| garbage dumps | Total Number of Seeds | CA (%) | SA (%) |
|---|---|---|---|
| 1 | 102 | 9 | 91 |
| 2 | 80 | 15 | 85 |
| 3 | 69 | 16 | 84 |
| 4 | 77 | 13 | 87 |
| 5 | 50 | 8 | 92 |
| 6 | 125 | 12 | 88 |
|  | 503 | $\overline{X} = 12,16 \pm DT = 2,91$ | $\overline{X} = 87,84 \pm DT = 2,91$ |

| Stores | Total Number of Seeds | CA (%) | SA (%) |
|---|---|---|---|
| 1 | 398 | 93 | 7 |

When a t - Student analysis for related measures is applied to the data (after a check for normality and homogeneity), a significantly higher proportion of seeds are obtained from which the aryl was removed in the garbage t = - 29.064; gl = 5; p <0.0001, which is in consequence with the inverse result found in their stores.

**7.1.11. What is the relative importance of ants on *Ulex parviflorus*?**
**1) Study of ants.**
The ants analyzed correspond only to two species: *Camponotus cruentatus* and *Messor barbarus*. The genus *Camponotus* is not granivorous nor is it suspected to be, but it is easy to find between *Ulex Parviflorus* and other scrubs and even trees, collecting sugary secretions from aphids. The genus *Messor* is a granivorous species, almost strictly of which a total of nine species are known for the Iberian Peninsula, including *Messor maroccanus*, *Messor hispanicus*, *Messor lusitanicus*, *Messor lobicornus*, *Messor celiae* and *Messor bouvieri*. However, the other species *Messor structor*, *Messor barbarus* and *Messor capitatus* have a wider distribution in Europe. (Tinaut et al, 1994; Bernard, 1968).

**2) Density of ants.**
The number of ants of the *Messor* species decreases in relation to the decrease in the number of fruits present in the plot, but the number of ants *Camponotus* is almost constant (Figures 34 and 35).

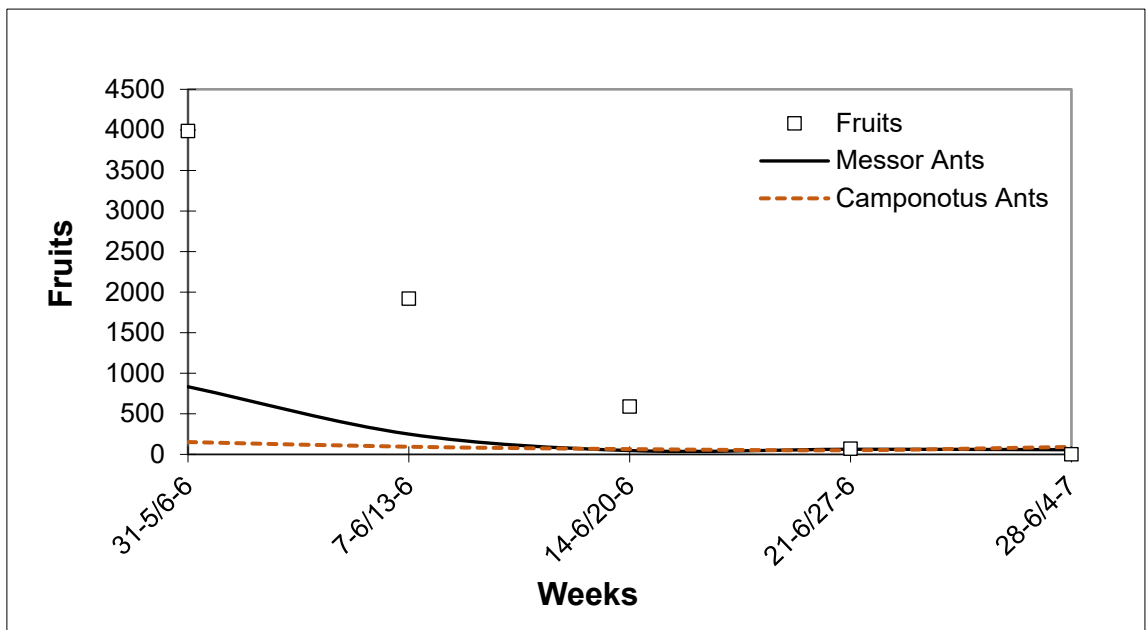

**Figure 34: Relationship: Ants Density/Seed Dispersion.**

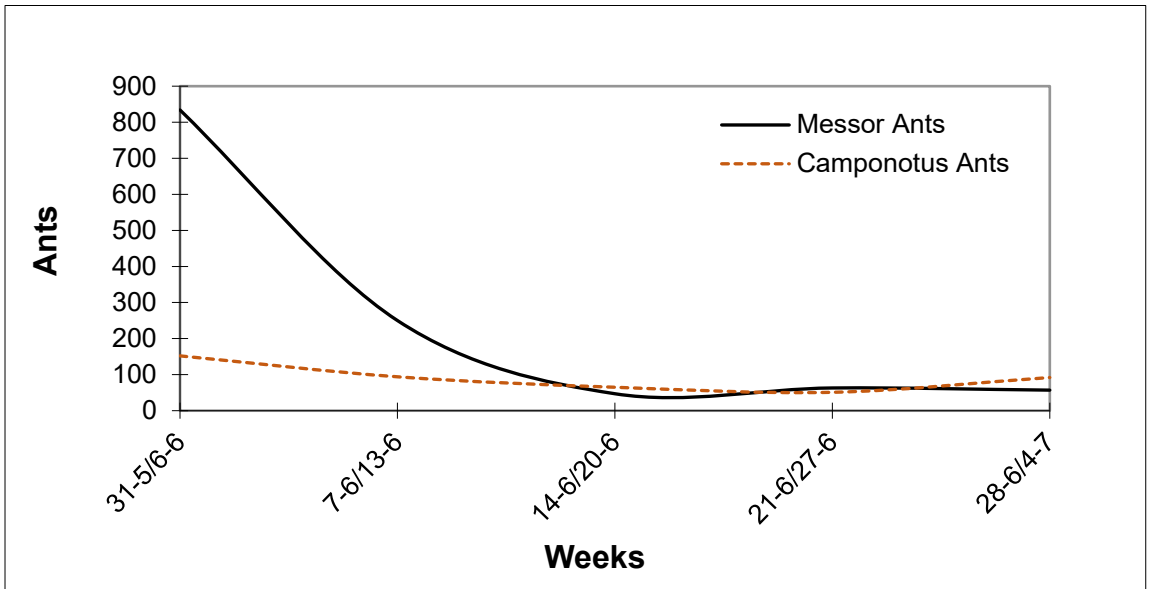

**Figure 35: Density of Ants on the plot**

**c) Presence of ants in the area.**
The anthills of *Messor barbarus* found were 6 (Figure 36), located at:

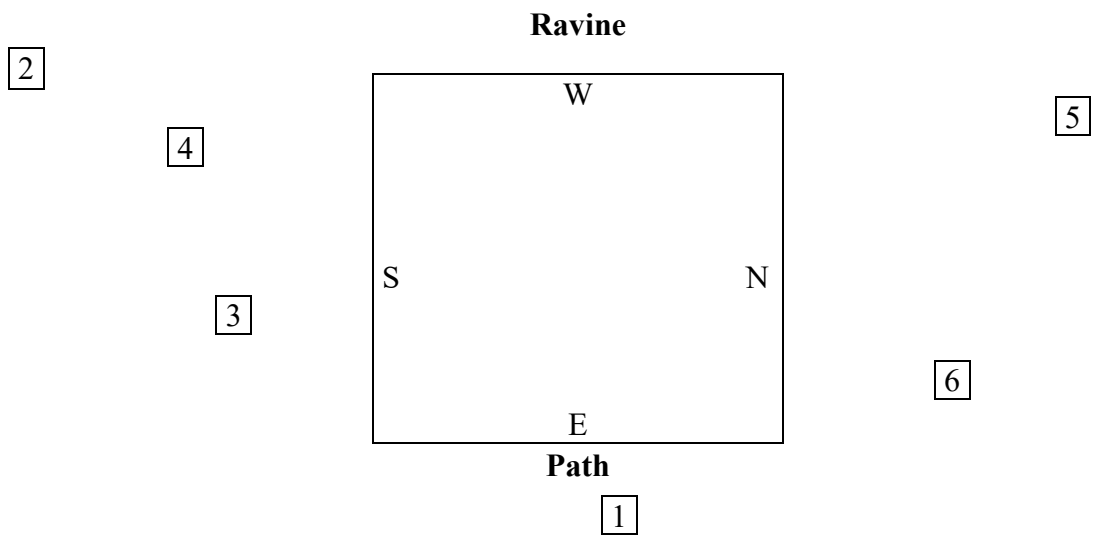

**Figure 36: Location of anthills.**

The density of anthills in the study area is 6/400 m2 which corresponds to approximately 0.01 anthills/m$^2$. Compared with other studies whose results correspond to 0.03 anthills / m2 (Acosta et al., 1992), they indicate that the impact of the ant *Messor barbarus* in the area is relatively significant.

## 8. DISCUSSIONS AND CONCLUSIONS

1) The flowering of this species is early and very numerous favoring the decrease of competitiveness from other species and effectiveness in the formation of seeds to be pollinated by the bees. *Ulex parviflorus* manages to save its difficulties because it competes in periods of pollination with *Anthyllis cytisoides* L., *Rosmarinus officinalis* L. and *Thymus vulgaris* L.

2) The study of the pollination of *Ulex parviflorus*, corroborates other studies carried out where it is also affirmed that the optimal period of pollination during the day occurs between 08:00 and 18:00. Thus the time available for pollination of any flower can be as much as 10 hours per day, for a maximum of 8 days, but pollination is most effective for 4 days (Richards, 1986). The effectiveness of species pollinated by bees is high. The order of this effectiveness varies little from year to year. Richards (1987) found studying the species *Astragalus cicer* L. that where pollination by bee predominated, *Apis mellifica* the percentage of seeds per pod was 5.2 (1978-1982 was 71% with a range of 67-77% ), when the individuals belonged to the species *Bombus sp.* and other native bees the percentage of seeds per pod was 4.6 (percentage of 66% with a range of 51-80%). Other authors have also found similar results, Zimmerman (1979) studying the species *Polemonium foliosissimum* finds that there is a positive correlation between the group of seeds that are obtained per flower with the number of visits of the bees that has received. In general, if there are several species visited by the same pollinator in the same area, the competition is strong enough to cause divergence in flowering periods (Gentry, 1974; Reader, 1975; Stiles, 1977). During the flowering period, the plants are much more attractive to pollinators acting as a selection advantage (Zimmerman$^a$, 1980). Early flowering contributes to increasing the number of seeds per flower (Zimmerman$^b$, 1980). Some authors have observed the presence of rare but highly attractive species for bees in associated plants (Beattie et al., 1973; Heinrich, 1976). This seems to be, in general, one to one, a correspondence between attraction and abundance; indicating that the abundance multiplied by the attraction is equal to a constant. This advantage underscores the conclusion that all species receive approximately the same number of visits (Pleasants, 1980). Climatic conditions affect pollination of bees, bearing in mind that storms trigger electromagnetic waves that travel hundreds of kilometers. Not only do they cause interference to the radio but they also affect bees. When they return to the hive they become irritable. They have detected electromagnetic waves or perceived changes in the electrostatic energy of the air. These electrical charges affect the behavior of the bee in the same way they influence our mood, so sunny days with low wind and low humidity are best for good pollination. The ratio of workers returning to the hive from each plant is usually six bees per minute. Bee colonies can respond within 15 minutes, shifting to a much more asymmetric pattern in which the ratio of worker bee visits to plants drops about 50% while the ratios of visits from other plants can maintain constant (Sherman et al., 2001). The queen bee breeds all year, but increases her activity in spring and summer causing the increase of the work in the beehives to transport the nectar and to feed the young. February and March are not a period with high activity of the bees.The bees usually choose a flower that is abundant in the area where they are traveling, thus saving time. If this flower is very abundant in the zone near the hive, probably almost all the bees from this hive will choose the same type of flower. The presence of the bees makes the pollen spread more easily, which means better development of the species between 30% and 40%. The flowering of this species is early and very numerous favoring the decrease of competitiveness from other species and the effectiveness in the formation of seeds to be pollinated by the bees. *Ulex parviflorus* manages to save its difficulties because it competes in the periods of pollination with *Anthyllis cytisoides* L., *Rosmarinus officinalis* L. and *Thymus vulgaris* L.

3) The growth of the plant coincides with the ripening of the fruits. After dispersion of the seeds, the plant enters a period of rest, beginning approximately with the summer until the middle of autumn, when the flowering begins again. The study carried out on the fruiting period of this species shows that the period of primary dispersion runs from the end of March to June, presenting a period of maximum dispersion in June. Coinciding with a rise in temperature and a drop in relative humidity. This raises the relation existing between the first dispersion of this species and the action of the colonies of *Messor barbarus*. In the fruiting period, two problems arise: one, not all the fruits formed give rise to mature seeds as indicated above and two, the seeds that reach the soil suffer two actions of the insects one of dispersion and another of predation. Two insects responsible for seed predation have been detected: *Sitona regensteinensis* and *Bruchidius lividimanus*, both of which are seminivorous on genistae. The first has been cited on *Cytisus scoparius*, *Cytisus purgans*, *Cytisus laburnum*, *Ulex europaeus* and *Ulex nanus*. Predation does not occur directly on the seed, but the insect lays eggs on the fruit when the fruit is immature and the larvae develop inside the fruit, feeding on the seed until reaching maturity. This same process, but produced by corculionids, parasitic hymenoptera and lepidopteran larvae, has been observed in *Ulex parviflorus* (Herrera, 1985).

4) The seeds of *Ulex parviflorus* are violently expelled from their fruits (explosive dehiscence) that fall to the ground by gravity. If only this dispersion existed, all seeds should be located near the parent plant, but this does not occur with the possibility of secondary dispersion (Beattie and Culver, 1981). Ants are responsible for taking these seeds from the soil surface and redistributing them (Sernander, 1906, Bresinsky, 1963, Beattie and Lyons 1975, Handel 1978, Handel et al., 1981, Westoby et al., 1982). Seeds with an external appendage, called an aryl or elaiosome (Sernander, 1906) have better dispersion by ants, acting as an additional dispersing agent (Bennett and Krebs, 1987; Beattie, 1985; Brew, O'Dowd and Rae, 1989, Ohkawara and Higashi, 1994, Beattie and Culver, 1982).

5) This establishes a double dispersion that can maintain the population, providing a double bank of seeds. Seeds that have been transported to ant mounds and manipulated by ants germinate better, indicating an effect of the aryl, that is, a greater preference for seeds with an aryl. This may suggest that the aryl may be a vehicle for seeds to reach a suitable site for gemination (Culver and Beattie, 1980) and another group of seeds, which is not picked up by ants, is incorporated into an adjacent seed bank and which can persist in the soil (Bülow-Olsen, 1984; Baeza, 2001), because ants gradually loose their interest in seeds in the soil due to the deterioration of the aryl (Anderson, 1983). These seed banks persist in the soil, in a state of dormancy or lethargy (Moreno et al., 1992; Baeza and Vallejo, 2006) and create a mechanism of seed dispersal over time (Venable and Brown, 1988).

6) The study carried out on the fruiting period of this species reflects that the period of primary dispersion coincides with an increase in temperature and a decrease in relative humidity, which is related to the explosion of legumes. This raises the relationship between the primary dispersion of this species and the redispersion by the action of the colonies of *Messor barbarus*.

7) The temperature range for the activity of the colonies of *Messor barbarus* is 33°C (from 3 to 36°C) (López, Serrano and Acosta, 1992). The strategy of finding food for this species is clearly sensitive to the density of the resource (López, Acosta and Serrano, 1993). This, along with the high seed turnover rate obtained in the predation experiments (> 60%), confirms that the time of greater seed availability of *Ulex parviflorus*, due to its natural dispersion, coincides with the period of greatest harvesting activity of the ants.

8) The results show that the ants look for the seeds, but they do not follow a pattern of distances and orientation with respect to the primary dispersion of the seeds of *Ulex parviflorus* in the soil. This means that the seeds of *Ulex parviflorus* are not a resource especially sought by *Messor barbarus* in comparison with other resources and that therefore their search is only conditioned by other sources of food, by the direction of the anthill, geographical conditions, etc. The trunk-trails of the granivorous ants among them, the *Messor barbarus* (L.), have been considered as devices to avoid competition between neighboring colonies, which give rise to an irregular distribution of the territory of search for food, around the anthill. (Acosta, López and Serrano, 1995).

9) There is a principle of optimal performance that minimizes the costs of finding food in terms of risk of struggle or difficulties imposed by the mobility of worker ants. The results coincide with the advantage of the short ways back to the anthill (Acosta, López and Serrano, 1993). On the other hand, the vegetation cover and its structure affect the activity of the ants (Rosengren and Pamilo, 1978, Whitford, 1978, Fernández and Rodríguez, 1982, Lorber, 1982). Ants collect some of the seeds present in the soil and move them to the anthills. The aril is used for food and to allow the ant to be able to seize the seeds while leaving the germ without suffering significant alterations. (Bennett and Krebs, 1987). The aryls of wooded plants can dry out fairly quickly when the fruit is opened. A seed with a dry aril has less attraction to ants and are often no longer dispersed by them. Thus, the rapid collection of the seeds by the ants, immediately after opening and/or exploding the fruit can prevent both processes: the drying of the aril and the predation of the seed. (Gorb and Gorb, 1999). The seeds of *Ulex Parviflorus* collected by ants are present in the anthills, both the warehouse and in the ants' dumps and our results show that in the first case (store) are raw, i.e., maintain the aryl, and therefore may be considered as reserves. While the seeds present in the dump show a very important and statistically significant tendency to be without an aryl. Germination tests show that aryl-free seeds germinate better. Therefore, the ants *Messor barbarus* facilitate this germination.

10) Whether seeds were processed by the ants or transferred to the dump, or otherwise being accidentally unearthed, they may be ready for germination. We do not know if there will be a recruitment of new individuals from *Ulex parviflorus*, but the few works that exist in this regard show that the influence of these in the studied plant communities is generally not detected (Acosta, López and Serrano, 1992). The importance of ants for this species is that they make their seeds after a primary dispersion undergo another secondary dispersion process. This secondary dispersion was pointed out by Moreno et al., (1992) in his work on the primary dispersion of the seeds of *Cytisus multiflorus*, when verifying the existence of the dispersion by explosive dehiscence; in which only 35% of the seeds fell under the plant, making the rest in its surroundings; some reaching distances greater than 3 meters. It also points out that the seeds could be predated as is the case of this species, enter 8.4 and 25%. Short-distance dispersion increases the probability of seed survival (Howe and Smallwood, 1982).

11) Some authors interpret this type of dispersion of seeds as an escape from predation as well as the transport of seeds by the ants, storing them later in the anthills. Although the dispersion of the granivorous ants as in the case of *Messor barbarus* corresponds to only 0.1% of the total. (Detrain and *Tasse, 2000). Therefore Messor barbarus facilitates a secondary dispersion to Ulex parviflorus*, as well as a better germination when eating the aryl, but they are not a special resource for these and its choice is conditioned by the density of the resource with respect to other sources of feeding. The position of the anthills and their density could condition the appearance of new *Ulex parviflorus* plants.